\documentstyle[11pt]{article}
\newcounter{subequation}[equation]

\makeatletter 
 
\def\thesubequation{\theequation\@alph\c@subequation}
\def\@subeqnnum{{\rm (\thesubequation)}}
\def\slabel#1{\@bsphack\if@filesw {\let\thepage\relax
   \xdef\@gtempa{\write\@auxout{\string
      \newlabel{#1}{{\thesubequation}{\thepage}}}}}\@gtempa
   \if@nobreak \ifvmode\nobreak\fi\fi\fi\@esphack}
\def\subeqnarray{\stepcounter{equation}
\let\@currentlabel=\theequation\global\c@subequation\@ne
\global\@eqnswtrue
\global\@eqcnt\z@\tabskip\@centering\let\\=\@subeqncr
$$\halign to \displaywidth\bgroup\@eqnsel\hskip\@centering
  $\displaystyle\tabskip\z@{##}$&\global\@eqcnt\@ne
  \hskip 2\arraycolsep \hfil${##}$\hfil
  &\global\@eqcnt\tw@ \hskip 2\arraycolsep
  $\displaystyle\tabskip\z@{##}$\hfil
   \tabskip\@centering&\llap{##}\tabskip\z@\cr}
\def\endsubeqnarray{\@@subeqncr\egroup
                     $$\global\@ignoretrue}
\def\@subeqncr{{\ifnum0=`}\fi\@ifstar{\global\@eqpen\@M
    \@ysubeqncr}{\global\@eqpen\interdisplaylinepenalty \@ysubeqncr}}
\def\@ysubeqncr{\@ifnextchar [{\@xsubeqncr}{\@xsubeqncr[\z@]}}
\def\@xsubeqncr[#1]{\ifnum0=`{\fi}\@@subeqncr
   \noalign{\penalty\@eqpen\vskip\jot\vskip #1\relax}}
\def\@@subeqncr{\let\@tempa\relax
    \ifcase\@eqcnt \def\@tempa{& & &}\or \def\@tempa{& &}
      \else \def\@tempa{&}\fi
     \@tempa \if@eqnsw\@subeqnnum\refstepcounter{subequation}\fi
     \global\@eqnswtrue\global\@eqcnt\z@\cr}
\let\@ssubeqncr=\@subeqncr
\@namedef{subeqnarray*}{\def\@subeqncr{\nonumber\@ssubeqncr}\subeqnarray}
\@namedef{endsubeqnarray*}{\global\advance\c@equation\m@ne%
                           \nonumber\endsubeqnarray}

\makeatletter
\@addtoreset{equation}{section}
\makeatother
\renewcommand{\theequation}{\thesection.\arabic{equation}}

\def\dalemb#1#2{{\vbox{\hrule height .#2pt
        \hbox{\vrule width.#2pt height#1pt \kern#1pt
                \vrule width.#2pt}
        \hrule height.#2pt}}}
\def\square{{\mathord{\dalemb{6.8}{7}\hbox{\hskip1pt}}}}

\def\half{{\textstyle{1\over2}}}
\let\a=\alpha \let\b=\beta \let\g=\gamma \let\d=\delta 
 \let\h=\eta \let\q=\theta  
 \let\m=\mu \let\n=\nu   \let\r=\rho
\let\s=\sigma   \let\f=\phi

\def\nn{\nonumber} \def\bd{\begin{document}} \def\ed{\end{document}}
\def\ds{\documentstyle} \let\fr=\frac \let\bl=\bigl \let\br=\bigr
\let\Br=\Bigr \let\Bl=\Bigl 
\let\bm=\bibitem
\let\na=\nabla
\let\pa=\partial \let\ov=\overline
\def\ie{{\it i.e.\ }} 
\newcommand{\be}{\begin{equation}} 
\newcommand{\ee}{\end{equation}} 
\def\ba{\begin{array}}
\def\ea{\end{array}}
\def\ft#1#2{{\textstyle{{\scriptstyle #1}\over {\scriptstyle #2}}}}
\def\fft#1#2{{#1 \over #2}}
\def\del{\partial}
\def\sst#1{{\scriptscriptstyle #1}}
\def\oneone{\rlap 1\mkern4mu{\rm l}}
\def\e7{E_{7(+7)}}
\def\td{\tilde}
\def\wtd{\widetilde}
\def\im{{\rm i}}
\def\bog{Bogomol'nyi\ }
\def\q{{\tilde q}}
\def\hast{{\hat\ast}}
\def\0{{\sst{(0)}}}
\def\1{{\sst{(1)}}}
\def\2{{\sst{(2)}}}
\def\3{{\sst{(3)}}}
\def\4{{\sst{(4)}}}
\def\5{{\sst{(5)}}}
\def\6{{\sst{(6)}}}
\def\7{{\sst{(7)}}}
\def\8{{\sst{(8)}}}
\def\n{{\sst{(n)}}}
\def\oo{{\"o}}
\def\hA{\hat{\cal A}}
\def\ns{{\sst {\rm NS}}}
\def\rr{{\sst {\rm RR}}}
\def\tH{{\widetilde H}}
\def\tB{{\widetilde B}}
\def\cA{{\cal A}}
\def\cF{{\cal F}}
\def\tF{{\wtd F}}

\def\Z{\rlap{\sf Z}\mkern3mu{\sf Z}}
\def\ep{{\epsilon}}
\def\IIA{{\rm IIA}}
\def\IIB{{\rm IIB}}
\def\ads{{\rm AdS}}
\def\R{\rlap{\rm I}\mkern3mu{\rm R}}
\def\vp{{\varphi}}
\def\ns{{\sst{\rm NS}}}
\def\rr{{\sst{\rm RR}}}
 \def\cF{{\cal F}} \def\cA{{\cal A}} \def\cB{{\cal
B}} \def\hA{{\hat{\cal A}}} \def\td{\tilde} \def\wtd{\widetilde}
\def\e{{\epsilon}} 
\def\Z{\rlap{\sf Z}\mkern3mu{\sf Z}}

\def\hhs{{\qquad}}
\def\hs{\,\,}
\def\Ka{{K\"ahler\ }}
\def\tK{{{\wtd K}}}
\def\bZ{{{\bar Z}}}
\def\bzeta{{{\bar \zeta}}}
\def\bb{{{\bar \b}}}
\def\CP{{C\!P}}
\def\HP{{H\!P}}

\newcommand{\ho}[1]{$\, ^{#1}$}
\newcommand{\hoch}[1]{$\, ^{#1}$}
\newcommand{\bea}{\begin{eqnarray}} 
\newcommand{\eea}{\end{eqnarray}} 
\newcommand{\ra}{\rightarrow}
\newcommand{\lra}{\longrightarrow}
\newcommand{\Lra}{\Leftrightarrow}
\newcommand{\ap}{\alpha^\prime}
\newcommand{\bp}{\tilde \beta^\prime}
\newcommand{\tr}{{\rm tr} }
\newcommand{\Tr}{{\rm Tr} } 
\newcommand{\NP}{Nucl. Phys. }
\newcommand{\tamphys}{\it Center for Theoretical Physics,
Texas A\&M University, College Station, TX 77843}
\newcommand{\upenn}{\it Dept. of Physics and Astronomy, 
University of Pennsylvania,
Philadelphia, PA 19104}

\newcommand{\auth}{P. Hoxha, R.R. Martinez-Acosta and  C.N. Pope}

\begin{document}
\begin{flushright}
\hfill{CTP TAMU-xx/00 \\ 
May 2000}\\
\hfill{\bf hep-th/0005172}\\
\end{flushright}


\begin{center}
{\large {\bf Kaluza-Klein Consistency, Killing Vectors and K\"ahler Spaces} 
} 

\vspace{20pt}

\auth

\vspace{10pt}
{\hoch{}\tamphys}

\vspace{30pt}

\underline{ABSTRACT}
\end{center}

    We make a detailed investigation of all spaces $Q_{n_1\cdots
n_N}^{q_1\cdots q_N}$ of the form of $U(1)$ bundles over arbitrary
products $\prod_i \CP^{n_i}$ of complex projective spaces, with
arbitrary winding numbers $q_i$ over each factor in the base. 
Special cases, including $Q_{11}^{11}$ (sometimes known as $T^{11}$),
$Q_{111}^{111}$ and $Q_{21}^{32}$, are relevant for compactifications
of type IIB and $D=11$ supergravity.  Remarkable ``conspiracies''
allow consistent Kaluza-Klein $S^5$, $S^4$ and $S^7$ sphere reductions
of these theories that retain all the Yang-Mills fields of the
isometry group in a massless truncation.  We prove that such
conspiracies do not occur for the reductions on the $Q_{n_1\cdots
n_N}^{q_1\cdots q_N}$ spaces, and that it is {\it inconsistent} to
make a massless truncation in which the non-abelian $SU(n_i+1)$
factors in their isometry groups are retained.  In the course of
proving this we derive many properties of the spaces $Q_{n_1\cdots
n_N}^{q_1\cdots q_N}$ of more general utility.  In particular, we show
that they always admit Einstein metrics, and that the spaces where
$q_i=(n_i+1)/\ell$ all admit two Killing spinors. We also obtain an
iterative construction for real metrics on $\CP^n$, and construct the
Killing vectors on $Q_{n_1\cdots n_N}^{q_1\cdots q_N}$ in terms of
scalar eigenfunctions on $\CP^{n_i}$.  We derive bounds that allow us
to prove that certain Killing-vector identities on spheres, necessary
for consistent Kaluza-Klein reductions, are never satisfied on
$Q_{n_1\cdots n_N}^{q_1\cdots q_N}$.

{\vfill\leftline{}\vfill
\vskip 10pt \footnoterule
{\footnotesize
        \hoch{}        Research supported in part by DOE 
grant DOE-FG03-95ER40917 \vskip -12pt}  \vskip  14pt
}

\pagebreak
\setcounter{page}{1}

\tableofcontents
\addtocontents{toc}{\protect\setcounter{tocdepth}{2}}
\newpage

\section{Introduction}

    In its original form Kaluza-Klein reduction was used for the
purpose of deriving a four-dimensional theory comprising gravity, a
$U(1)$ gauge field and a dilatonic scalar, starting from pure gravity
in five dimensions.  The extra dimension is taken to be a circle, and
the five-dimensional metric is then assumed to be independent of the
coordinate $y$ on the circle.  Such a truncation is consistent, and
gives rise to an Einstein-Maxwell theory in $D=4$, coupled to the
dilatonic scalar field.  The consistency of the truncation is assured
because the reduction ansatz retains all the four-dimensional fields
that are independent of $y$, while setting all fields that would be
associated with $y$-dependent harmonics on $S^1$ to zero.  In a
similar vein, Kaluza-Klein reductions involving higher-dimensional
theories compactified on tori can also be considered, and again
consistent truncations where all fields are taken to be independent of
the torus coordinates can be performed.

   The situation is much less clear-cut in the case where one performs
a reduction on a curved internal manifold, such as a sphere.  The new
complication in such a case is that the harmonics on the internal
space associated with the massless fields in the lower dimension
typically now depend on the coordinates of the internal space.  This
causes no difficulty in a linearised analysis of small fluctuations
around a ground-state solution (see, for example, \cite{DNP}, and
references therein), but as soon as one wants to consider the full
non-linear structure of the theory it raises the possibility of
inconsistencies in a truncation to the massless sector.  In fact this
is more than a possibility; in general, there will definitely be
inconsistencies.  This makes it all the more remarkable that there
exist certain exceptional cases in which a fully non-linear sphere
reduction and truncation {\it is} completely and rigorously
consistent.  Many of the known cases involve special reductions of
supergravity theories, notably involving $S^7$ \cite{dwn} or $S^4$
\cite{vann1,vann2} reductions of $D=11$, the $S^5$ reduction of type
IIB,\footnote{The consistency of the $S^5$ reduction to
five-dimensional maximal gauged supergravity remains conjectural at
this time, but strong supporting evidence has been obtained, including
various explicit consistent reductions to subsets of the maximal
supergravity \cite{ten,d5gauge,clpst}, and an explicit expression for
the complete metric reduction ansatz \cite{piwa}.}  and a local $S^4$
reduction of the massive type IIA theory \cite{d6gauge}.  Other
exceptional examples of consistent sphere reductions in which all the
Yang-Mills gauge fields can be retained have also been found recently,
for cases that do not necessarily have any connection with
supersymmetry.  These comprise the reduction of the low-energy limit
of the bosonic string, in an arbitrary dimension $D$, on the 3-sphere
or the $(D-3)$-sphere, and the reduction of certain theories of
gravity plus a dilaton and a 2-form field strength in $D$ dimensions
on a 2-sphere \cite{clpnew}.  In all these cases, there is no known
group-theoretic proof for why the reduction should be
consistent.\footnote{A group-theoretic argument has been used in
\cite{clpst,clpnew} in order to prove that an $n$-sphere reduction of
a theory of gravity plus dilaton plus $n$-form field strength that
retained all the $SO(n+1)$ Yang-Mills fields in a massless truncation
could not be consistent except in the exceptional cases listed above.}

    Two approaches to proving the consistency of these supergravity
sphere reductions have been pursued in the literature.  For the $S^7$
\cite{dwn} and $S^4$ \cite{vann1,vann2} reductions from $D=11$, the
truncations to the maximally supersymmetric gauged $SO(8)$ and $SO(5)$
supergravities in $D=4$ and $D=7$ have been argued to be consistent by
demonstrating that consistent supersymmetry transformation rules in
the lower dimension can be extracted from the original ones in $D=11$.
A complete and rigorous proof of consistency along these lines would
in principle require the analysis of the supersymmetry transformation
rules to all orders, including quartic fermion terms, and the
difficulties in doing this are considerable.  However, it seems
reasonable to conclude that the already highly non-trivial success at
the quadratic level would persist to all orders.  The approach has
been used for the $S^7$ \cite{dwn} and $S^4$ \cite{vann1,vann2}
reductions, and in the latter case has allowed an explicit
construction of the exact bosonic reduction ansatz.  No analogous
complete results have been obtained for the $S^5$ reduction of type IIB
supergravity, but it seems highly likely to be consistent also.

   The alternative approach to proving the consistency of a
Kaluza-Klein reduction is a more direct one, in which one explicitly
constructs a reduction ansatz which, when substituted into the full
set of higher-dimensional equations of motion, gives a consistent
embedding provided that the lower-dimensional equations of motion are
satisfied.  This approach has been used to provide a complete proof of
the consistency in several sphere reductions, where further
truncations to subsets of the fields of the maximal massless
supermultiplet are made.  Cases that have been fully proven by this
means include $N=2$ gauged $SU(2)$ supergravity in $D=7$ by an $S^4$
reduction from $D=11$ \cite{d7gauge}; the $N=4$ gauged $SU(2)\times
U(1)$ supergravity in $D=5$ by an $S^5$ reduction from type IIB in
$D=10$ \cite{d5gauge}; the $N=4$ gauged $SO(4)$ supergravity by $S^7$
reduction from $D=11$ \cite{d4gauge}, and the $N=2$ gauged $SU(2)$
supergravity in $D=6$ by a local $S^4$ reduction from the massive type
IIA theory in $D=10$ \cite{d6gauge}.  (In this last example $N=2$ is
in fact the largest supersymmetry for gauged supergravity in $D=6$,
even though ungauged $N=4$ supergravity exists.)  In addition, the
consistency of the truncations of the $S^4$, $S^5$ and $S^7$
reductions to include gravity and all the diagonal scalars of the
$SL(5,R)/SO(5)$, $SL(6,R)/SO(6)$ and $SL(8,R)/SO(8)$ submanifolds of
the full scalar cosets of the maximal supergravities \cite{cglp} 
have been fully demonstrated \cite{clps}.

    It is significant that all these examples involve reductions on
spheres.  At the linearised level there is no reason why one should
not consider also reductions on internal spaces of other topologies.
Examples that have been considered in the past include the Einstein
spaces contructed as $U(1)$ bundles over $CP^2\times S^2$ and
$S^2times S^2\times S^2$, as compactifications of eleven-dimensionsal
supergravity, and $U(1)$ bundles over $S^2\times S^2$, as
compactifications of type IIB supergravity.  The first two examples
were first discussed in detail in \cite{fre1,fre2}, where they were
constructed as the coset spaces $M^{pqr}=SU(3)\times SU(2)\times
U(1)/(SU(2)\times U(1)\times U(1)$ and $Q^{pqr}=SU(2)^3/(U(1)\times
U(1))$ respectively, and the linearised massless spectra were
obtained.  They were subsequently reconstructed from the viewpoint of
$U(1)$ bundles over $CP^2\times S^2$ and $S^2\times S^2\times S^2$
respectively, in \cite{page1,page2}, where a stability analysis was
also given.  The complete massive spectrum was obtained in a
linearised analysis in \cite{fre3,fre4,fab1,fab2}.  The five-dimensional
example, the $U(1)$ bundle over $S^2\times S^2$, was discussed from
the AdS/CFT viewpoint in \cite{klebwit}, and in a field-theoretic
context in \cite{dufflupope}.  The full Klauza-Klein spectrum was
obtained in \cite{ceres}, and its matching with the conformal
operators of the dual CFT was obtained.

    Amongst the lower-dimensional massless fields that would result
from reductions such as these will be Yang-Mills gauge bosons with gauge group
given by the isometry group of the internal space.  One may wonder
whether a consistent truncation that includes the Yang-Mills gauge
fields is possible in these more general reductions too, or whether it
is a special feature of the spherical spaces that ensures the
consistency.  

    Some results on certain of these more general reductions were
obtained in previous studies.  In this paper, we shall address the
question in a slightly broader context.  The conclusions will be
similar to those reached in the previous cases, namely that in general
the reductions on non-spherical internal spaces do not allow
consistent truncations to the massless sector, even in those
exceptional and remarkable theories where consistent sphere reductions
are possible.  In fact it is much easier to demonstrate the
inconsistency of an inconsistent truncation than to prove the
consistency of a consistent one.  As was discussed in \cite{zilch},
when there are inconsistencies they tend to show up in relatively
easily-studied sectors of the theory, at the level of cubic
interaction terms in the Lagrangian.  In this paper we shall be
considering a specific type of cubic interaction, namely terms that
are bilinear in the lower-dimensional gauge fields, and that couple to
lower-dimensional linearised spin-2 fields.  This sector provides a
necessary condition for consistency of a truncation; in general it
turns out that the bilinears in gauge fields can act as sources for
{\it massive} as well as massless spin-2 excitations.  If this
happens, then setting the massive spin-2 fields to zero is
inconsistent with the higher-dimensional equations of motion, and so
the reduction is established to be an inconsistent one.  We derive
this condition in section 2.

    As we shall discuss, the absence or presence of these kinds of
trilinear couplings is governed by whether or not the Killing vectors
on the internal space satisfy a certain quadratic identity.  We shall
show that although the full set of $SO(n+1)$ Killing vectors on the
sphere $S^n$ do indeed satisfy the identity, implying no inconsistency
in this sector, the full sets of Killing vectors in the case of other
internal manifolds do not.  In particular, we shall show by this means
that for the 5-dimensional space $Q(1,1)$ (sometimes called $T^{11}$),
which can be described as a $U(1)$ bundle over $S^2\times S^2$, only
the Killing vector of the $U(1)$ factor in its $U(1)\times SU(2)\times
SU(2)$ isometry group satisfies the consistency condition.  Thus in a
reduction of type IIB supergravity on the $Q(1,1)$ space, only the
$U(1)$ gauge field of the $N=2$ supergravity multiplet can be
consistently retained in a massless truncation, whilst the $SU2)\times
SU(2)$ gauge fields of the matter multiplets must be set to zero.

    In order to demonstrate that the $SU(2)\times SU(2)$ Killing
vectors of the $Q(1,1)$ space fail to satisfy the consistency
criterion, it is helpful to obtain an explicit construction for them.
Motivated by this, we have undertaken a rather more general
investigation of the construction of Killing vectors in spaces of this
kind.  The base space $S^2\times S^2$ in the construction of $Q(1,1)$
as a $U(1)$ bundle is \Ka, and in fact in this specific case it itself
is an Einstein space.  More generally, one can consider the $U(1)$
bundle spaces over any Einstein-\Ka base space, or over a product of
Einstein-\Ka spaces.  Other relevant examples of this kind are the
7-dimensional $M(3,2)$ and $Q(1,1,1)$ spaces that have been used in
compactifications of $D=11$ supergravity \cite{fre1,fre2}.  These
arise, respectively, as $U(1)$ bundles over $\CP^2\times S^2$ and over
$S^2\times S^2 \times S^2$ \cite{page1,page2}.  In all the cases, the
curvature of the $U(1)$ connection is proportional to the sum of the
\Ka forms on the factors in the base space.

    Intuitively, one expects that if the base space has an isometry
group $G$, and the curvature of the $U(1)$ bundle is invariant under
$G$, then the isometry group of the bundle space should be at least
$U(1)\times G$.  In section 3 we show how to make this precise, and we
obtain explicit formulae that allow one to ``lift'' the Killing
vectors of the base space to Killing vectors in the bundle space.  The
situation is especially nice if the base space is Einstein-\Ka, or
else a product of Einstein-\Ka spaces, and we show how one can then
express the Killing vectors of the base, and hence of the bundle
space, in terms of certain scalar harmonics on the Einstein-\Ka
factors in the base space.

    In section 4 we specialise the discussion to the case where the
Einstein-\Ka manifolds $M_i$ in the product base space $M=M_1\times
M_2\times M_N$ are taken to be $M_i = \CP^{n_i}$.  We denote the
corresponding bundle spaces by $Q_{n_1 n_2\cdots n_N}^{q_1 q_2 \cdots
q_N}$, where $q_i$ is the winding number of the $U(1)$ fibre over the
factor $M_i=\CP^{n_i}$ in the product base manifold. The three examples
$Q(1,1)$, $M(3,2)$ and $Q(1,1,1)$ described above are special cases
within this general class, namely $Q_{11}^{11}$, $Q_{21}^{32}$ and
$Q_{111}^{111}$ respectively.  We give an explicit construction of the
$SU(n+1)$ Killing vectors of $\CP^n$ in terms of certain scalar
harmonics.  Using this construction and the results from section 3, we
are able to lift all the Killing vectors of the product base manifold
for  $Q_{n_1 n_2\cdots n_N}^{q_1 q_2 \cdots q_N}$ into the total
bundle space, thereby exhibiting its $U(1)\times \prod_i SU(n_i+1)$
isometry group.  

      We prove also that all the $U(1)$ bundle spaces $Q_{n_1
n_2\cdots n_N}^{q_1 q_2 \cdots q_N}$ admit Einstein metrics of
positive Ricci curvature, for all possible choices of the winding
numbers $q_i$, provided only that they do not all vanish.  In addition
we show that when all the $q_i$ are given by $q_i=n_i+1$, the Einstein
metric admits 2 Killing spinors.

     In section 5 we make use of some of the general results from
sections 3 and 4, to show explicitly that the Killing vectors in the
$SU(n_i+1)$ factors in spaces such $Q_{11}^{11}$, $Q_{21}^{32}$ and
$Q_{111}^{111}$ do not satisfy the consistency criterion for the
Kaluza-Klein reductions.  These results support the suggestion, made
in \cite{pope1}, that only the massless fields in the {\it
supergravity} multiplet, as opposed to any massless {\it matter}
multiplets, can be consistently retained in a Kaluza-Klein reduction
using a curved internal space.  Thus the reason why spheres work so
well in Kaluza-Klein supergravity reductions is because they maximise
the number of Killing spinors, and thus their supergravity multiplets
are larger than those for any other choice of compactifying space.
More generally, in section 5, we analyse the analogue of the
consistency condition for bundle spaces of arbitrary dimension, and we
show that always the Killing vectors associated with the isometries of
the base manifold, when it is a product of two or more complex
projective spaces, do not satisfy the consistency condition.  Two
appendices contain some further general results, including an
iterative construction of real metrics on $\CP^n$, and a detailed
analysis of certain bounds on integrals involving the scalar
eigenfunctions on $CP^n$, which are needed for the results in section 5.

\section{Consistency conditions on Killing vectors in Kaluza-Klein
reductions}

    In this section, we shall focus principally on the Kaluza-Klein
reduction of type IIB supergravity on a 5-dimensional internal space
${\cal M}_5$.  Analogous results have previously been obtained for
reductions of $D=11$ supergravity \cite{dnpw}, 
and we shall mention these briefly at the end of the section.
 
    Since our goal will be to derive a {\it necessary}
condition for the consistency of the reduction, with a view to showing
that the condition is not in fact satisfied except in very special
circumstances, it will be sufficient to carry out an analysis that is
based on a linearised approximation.  Thus we shall consider a
situation where ${\cal M}_5$ is an Einstein space of positive Ricci
curvature, and we shall consider small fluctuations around the
AdS$_5\times {\cal M}_5$ Freund-Rubin background.  In particular, we
shall consider the Yang-Mills gauge bosons associated with the
isometry group $G$ of the internal space ${\cal M}_5$.  

    Although we shall consider only the linearised ansatz for the
gauge bosons this will actually enable us to consider the effects of
non-linear terms in theory, and in particular to show that bilinears
in the gauge fields will in general act as sources for massive spin-2
fields.  The reason why we can use a linearised ansatz for this
purpose is that gauge invariance ensures that there can be no
additional contributions from a full non-linear reduction ansatz that
could ``help out'' and resolve the consistency problems that we shall
be able to reveal.  Thus, since our goal is only to prove
inconsistency, not consistency, the analysis presented here will be
sufficient.\footnote{Note that we shall ignore the contributions of
other five-dimensional fields, including the scalar fields, in this
discussion.  Truncating out these fields, while keeping the Yang-Mills
gauge fields, is itself an inconsistent procedure, since the
Yang-Mills fields would in principle act as sources for them.  The
point is, though, that these are quite distinct and separate
inconsistencies, which would show up in different sectors of the
theory.  By focusing, as we shall, on the five-dimensional spacetime
components of the ten-dimensional Einstein equation we shall be able
to isolate a particular inconsistency that is independent of the
neglect of the other fields.  In other words, including the other
fields in the ansatz would not help to resolve the inconsistency that
we shall exhibit.}

   The fields of the type IIB theory that are relevant for this
discussion are the metric tensor $\hat G_{MN}$ and the self-dual
5-form field strength $\hat H_5$, which we may write as $\hat H_5 =
\hat G_5 + {\hat *\hat G_5}$.  
The type IIB equations of motion for these fields are then
\bea
{\hat R}_{MN} &=& \frac {1}{96} {\hat H}_{MPQRS} {\hat H}_N{}^{PQRS}\,,\nn\\
d{\hat H}_5 &=& d*{\hat H}_5=0\,.\label{eqsofmotion}
\eea
The Freund-Rubin AdS$_5\times {\cal M}_5$ ground-state solution is
then obtained by setting $\hat G_5= 4m\, \ep_5$, where $\ep_5$
is the spacetime volume form and $m$ is a constant.  The equations of
motion are then satisfied if the Ricci tensors in the five-dimensional
spacetime and the internal space ${\cal M}_5$ satisfy
\be
R_{\mu\nu}= -4 m^2\, g_{\mu\nu}\,,\qquad {\rm and}\qquad 
R_{mn} = 4 m^2\, g_{mn}
\ee
respectively.  Thus we may take the spacetime metric to be AdS$_5$
with cosmological constant $-4m^2$, and ${\cal M}_5$ to be any
5-dimensional Einstein space with positive cosmological 
constant\footnote{We shall adopt the convention throughout this paper
of referring to the constant of proportionality $\Lambda$ in the
relation $R_{ab} = \Lambda\, g_{ab}$ on an Einstein space as the
cosmological constant.  It sometimes differs by a dimension-dependent
factor from other terminologies in the literature, but this one has the
merit of simplicity.} $4m^2$.

     We may now consider the contributions of the five-dimensional 
Yang-Mills gauged bosons in the ans\"atze for the ten-dimensional
metric and 5-form field strength, at the leading-order linearised
level.  For the metric, this will be
\be
d{\hat s}^2=e^{\a} e^{\b}\h_{\a\b}+(e^a-K^{Ia}A^I)(e^b-K^{Jb}A^J)
\,\d_{ab},
\ee
where $e^{\a}=e^{\a}(x)$ is the vielbein in the $d=5$ spacetime,
$e^a=e^a(y)$ is the vielbein in the internal space,
$K^{Ia}=K^{Ia}(y)=K^{Im}(y)\, {e_m}^a(y)$ are the orthonormal
components of the Killing vectors which generate the isometry group
$G$ of the internal space ${\cal M}_5$, and
$A^I=A^I(x)=e^{\a}(x)A^I_{\a}(x)$ are the Yang-Mills vector potentials
of the Kaluza-Klein reduction. The Killing vectors satisfy
\be
[K^I,K^J]={f^{IJ}}_K\,  K^K,
\ee
where ${f^{IJ}}_K$ are the structure constants of $G$. The Yang-Mills field
strengths $F^I={\half}F^I_{\a\b}e^{\a} \wedge e^{\b}$ are given by
\be
F^I=dA^I+{\half}f^{IJK}A^J \wedge A^K.
\ee
In an orthonormal basis ${\hat e}^A$ for $d{\hat s}^2$ we find that the Ricci
tensor given by
\bea
{\hat R}_{\a\b} &=& R_{\a\b}-\half K^{Ia} {K^J}_a {F^I}_{\a\g}
{{F^J}_{\b}}^{\g}\,,\nn\\
{\hat R}_{ab} &=& R_{ab}+ {\frac{1}{4}} {K^I}_a {K^J}_b {F^I}_{\a\b}
F^{J\a\b}\,,\nn\\
{\hat R}_{\a b} &=& {\hat R}_{b\a}=-\half {K^I}_b
(D_{\b}{{F^I}_{\a}}^{\b}),\label{ricci}
\eea
where $D_{\g}$ is the Yang-Mills gauge-covariant derivative,
\be
D_{\g}{F^I}_{\a\b}=\nabla_{\g}\,{F^I}_{\a\b}+f^{IJK}{A^I}_{\g}{F^K}_{\a\b}.
\ee
The curvature scalar is
\be
\hat R=R_{(5)}+R_{(M)}-{\frac{1}{4}} {K^I}_a K^{Ja} {F^I}_{\a\b}
F^{J\a\b},\label{curv-scalar}
\ee
where $R_{(5)}$ and $R_{(M)}$ are the curvature scalars in spacetime and
the internal space $M$ respectively.

   The gauge fields also enter in the linearised ansatz for the 5-form
field strength \cite{KRV}, as follows:
\be
{\hat G}_5=4m\, \ep_5-\frac{1}{m}*F^I \wedge dK^I.\label{G5}
\ee
Substituting (\ref{ricci}), (\ref{curv-scalar}) and ({\ref{G5}) into
(\ref{eqsofmotion}) we find that the five-dimensional spacetime
components of the ten-dimensional Einstein equation in (\ref{eqsofmotion}) 
are given by
\be
R_{\m \nu}-\half g_{\m \nu}(R_{(5)}+R_{(M)}) = \half ({F^I}_{\m\r}
{F^J}_\nu^\r-\fft14  g_{\m\nu}{F^I}_{\s\r}{F^J}^{\s\r}) \,
 Y^{IJ},\label{d5eom} 
\ee
where
\be
 Y^{IJ}=Y(K^I,K^J)\equiv {K^I}_m K^{Jm}+\frac{1}{2m^2} \nabla_m{K^I}_n 
\nabla^m K^{Jn}.\label{kappa}
\ee

    The possibility of an inconsistency in the Kaluza-Klein reduction
becomes apparent from equation (\ref{d5eom}). The left-hand side is
independent of the coordinates $y$ on the internal space ${\cal M}_5$,
while the right-hand side is in general $y$-dependent, since the
Killing appearing in $ Y^{IJ}$ are in general $y$-dependent.  If the
right-hand side does have $y$-dependence then this is an indication
that the assumption that only the {\it massless} spin-2 field (the
five-dimensional spacetime metric) could be retained in the truncation
is an invalid one.  One can interpret any $y$-dependence on the
right-hand side as indicating that there are bilinear terms,
built from the Yang-Mills field strengths, that would act as sources
for massive spin-2 fields.  Thus it would be inconsistent to make a
truncation where the massless gauge bosons are retained, while the
massive spin-2 fields are set to zero.
 
    By contrast, this inconsistency problem would be evaded if the
quantity $Y_{IJ}$ defined in (\ref{kappa}) happened to be independent
of $y$.  In such a case one could, by taking appropriate
linear combinations of the Killing vectors, arrange that
\be
 Y^{IJ}=\b \, \d^{IJ}\,,\label{condition}
\ee
where $\b$ is a constant.  In this circumstance, (\ref{d5eom}) would
become precisely the desired five-dimensional Einstein equation, with
the right-hand side being the energy-momentum tensor of the
Yang-Mills fields.

    Remarkably, all the Killing vectors on the round 5-sphere {\it do}
satisfy the condition (\ref{condition}), thus providing strong
evidence for the probable consistency of the $S^5$ reduction of type
IIB supergravity.  On the other hand, it seems that for any other
Einstein space ${\cal M}_5$ with positive cosmological constant, the
Killing vectors do not in general satisfy the condition
(\ref{condition}), and thus in these cases a consistent massless
truncation in which all the Yang-Mills gauge bosons are retained is
not possible.

   Note that there is an analogous criterion for the consistency of
reductions of $D=11$ supergravity.  This was derived for
compactifications on seven-dimensional Einstein spaces in \cite{dnpw},
and takes the identical form (\ref{kappa}) where the Ricci tensor on
the internal seven-dimensional space is given by $R_{mn} = 6 m^2\,
g_{mn}$.  Similarly, the analogous consistency condition will arise
for reductions of $D=11$ supergravity on four-dimensional internal
spaces, and indeed for any of the cases where consistent sphere
reductions are known to be possible.  A detailed enumeration of these
cases is given in \cite{clpnew}.  In fact in general one can show that
for any round sphere $S^n$ with $R_{mn} = (n-1)\, m^2 \, g_{mn}$, the
condition (\ref{condition}) is satisfied by all the $SO(n+1)$ Killing
vectors.  (This does not necessarily mean that a consistnet
Kaluza-Klein reduction on $S^n$ is possible, though.)

    It is worth pausing here to emphasise that although we have
derived the consistency condition that (\ref{kappa}) must be constant
by means of a consideration only of the {\it linearised} ansatz for
the Kaluza-Klein reduction of the gauged fields, the result is a
completely general one.  The reason for this is discussed in detail in
\cite{zilch}).  The crucial point is the following.  If (\ref{kappa})
is $y$-dependent, this shows that in a complete Kaluza-Klein reduction
in which all the massive as well as massless fields were retained,
there would be trilinear couplings involving one power of a heavy
spin-2 field, say $H^{IJ}_{\mu\nu}$, coupling to the bilinear source term 
quadratic in gauge fields
$F^I_{\m\r}$ on the right-hand side of (\ref{d5eom}):
\be
{\cal L}_{\rm int} = H^{IJ\mu\nu} \,  ({F^I}_{\mu\r}
{F^J}_\nu^\r- \fft14 g_{\m\nu}{F^I}_{\s\r}{F^J}^{\s\r})
\,.\label{interaction} 
\ee
Now, the masses of all the lower-dimensional massive fields are
acquired through a Higgs mechanism, and so it follows that the
original gauge invariances must remain unbroken.  In consequence, the
lower-dimensional massive spin-2 field $H^{IJ}_{\mu\nu}$ must have a
gauge invariance, implying that the source-current that couples to it
must be conserved \cite{zilch}.  Indeed, from an order-by-order
analysis it follows that the bilinear current must be conserved by
virtue of the free field equations.  The bilinear current
\be
 ({F^I}_{\mu\r}
{F^J}_\nu^\r-\frac{1}{4} g_{\m\nu}{F^I}_{\s\r}{F^J}^{\s\r})
\ee
appearing in (\ref{interaction}) is the {\it unique} one with this
property, and so it is not possible for it to receive any corrections
as a result of including higher-order non-linear terms in the
reduction ansatz.  Thus there is no possibility that the inconsistency
we are highlighting could ``disappear'' in a more complete
higher-order analysis.  If the quantity (\ref{kappa}) turns out to be
$y$-dependent, then no consistent Kaluza-Klein reduction in which the
associated gauge fields are retained is possible.  

    In order to proceed with our discussion, it is now necessary to
study in detail the Killing vectors on the internal space.  We shall
give a rather general discussion, which encompasses many of the
compactifications of type IIB supergravity and eleven-dimensional
supergravity as special cases.  Later, in section 5, we shall apply
these results to study the consistency of the Kaluza-Klein reductions.

\section{Construction of Killing vectors on the internal space}

\subsection{Killing vectors on $U(1)$ bundles}

    Consider a $D$-dimensional manifold with a group $G$ of isometries.
Suppose that there exists a $U(1)$ connection on $M$ whose curvature
is invariant under the isometry group $G$.  One then expects that the
natural metric on the $(D+1)$-dimensional bundle space with $U(1)$
fibers corresponding to this $G$-invariant $U(1)$ connection should
contain $G\times U(1)$ in its isometry group.\footnote{Generically,
this will be the full isometry group of the bundle space, but in
special cases it could be a larger group containing $G\times U(1)$ as
a subgroup.  An example is when the base manifold is $\CP^n$, and the
bundle space is the sphere $S^{2n+1}$.  If the length of the $U(1)$
fibres is chosen so as to give the ``round'' sphere, then the generic
$SU(n+1)\times U(1)$ isometry group in the bundle enlarges to $SO(2n+2)$.}  

    To see that this is indeed the case, suppose that the metric on
the base manifold is $ds^2$, and that the invariant $U(1)$ connection
is $A$.  The natural metric on the $(n+1)$-dimensional bundle space is
then taken to be
\be
d\hat{s}^2 = c^2 (dz - A)^2 + ds^2\,,\label{bundlemet} 
\ee
where $c$ is a constant, and $z$ is the coordinate on the $U(1)$
fibre.  (From this point on, we adopt the convention that
quantities with hats refer to the total bundle space $\hat M$,
while quantities without hats refer to the base space $M$.  We shall
use indices $m,n,\ldots$ in the total bundle space $\hat M$, and
indices $a,b,\ldots$ in the base space $M$.)
In the obvious orthonormal frame, the Riemann tensor for $d\hat
s^2$ has components
\bea
\hat R_{abcd} &=& R_{abcd} -\ft14 c^2\,
(F_{ac}\, F_{bd} - F_{ad}\, F_{bc} + 2 F_{ab}\, F_{cd})\,,\nn\\
\hat R_{zazb} &=& \ft14 c^2\, F_{a}{}^c\, F_{bc}\,,\label{riemman2}\\
\hat R_{abcz} &=& \ft12 c \,\nabla_c \,F_{ab}\,,\nn
\eea
where $R_{abcd}$ is the Riemann tensor of the metric $ds^2$ on the
base manifold.

    From (\ref{riemman2}) it follows that the components of the Ricci
tensor for $d\hat s^2$ are
\bea
\hat R_{ab} &=& R_{ab} - \ft12 c^2\, F_{ac}\, F_b{}^c\,,\nn\\
\hat R_{zz} &=& \ft14 c^2\, F^{ab}\, F_{ab}\,,\label{ricci2}\\
\hat R_{a z} &=& -\ft12 c\, \nabla^b\, F_{ab}\,,\nn
\eea
where $R_{ab}$ is the Ricci tensor on the base manifold.

    We now make the following ansatz in order to lift a 
Killing vector $K$ on the base manifold $M$ to a Killing vector $\hat
K$ on the total bundle space $\hat M$:
\be 
\hat{K} = K + h\, \partial_{z}\,, \label{kv_bundle}
\ee
where $h$ is a function to be determined. By
substituting our ansatz (\ref{kv_bundle}) into the Killing equation of
the bundle space:
\be
\hat{\nabla}_m\, \hat{K}_n + \hat{\nabla}_n\, \hat{K}_m=0\,,\label{kveq_bundle}
\ee
we find that $\hat{K}$ is a Killing vector on the bundle space
provided that  $h$ satisfies the following two equations:
\bea
\partial_{a}h &=& {\cal L}_K\, A_a \,,\label{lie}\\
\partial_{z}h &=& 0 \,,\label{zcon}
\eea
where ${\cal L}_K\, A_a$ is the Lie derivative, defined by
\be
 {\cal L}_K\, A_a \equiv  K^b\, \del_b\, A_a + A_b\, \del_a\, K^b =
 K^b\, \nabla_b\, A_a + A_b\, \nabla_a\, K^b\,.
\ee

   Equation (\ref{lie}) can be rewritten in terms of the field
strength $F=dA$ as
\be
\partial_{a}h = 
K^b\,F_{ba} + \partial_{a} (K^b \, A_b)\,.\label{u1_cond1}
\ee
It is easy to see that the two equations (\ref{u1_cond1}) and
(\ref{zcon}) always admit a solution, provided that $F$ is invariant
under the action of the Killing symmetry generated by $K$.  Clearly
(\ref{zcon}) is nothing more than the statement that $h$ is
independent of the fibre coordinate $z$.  The integrability condition
for solving (\ref{u1_cond1}) for $h$ is that the right-hand
side should be expressible as the gradient of a scalar.  Since the
second term is already a gradient, this means that we must just 
show that $\nabla_c(K^b\, F_{ba}) - \nabla_a(K^b\, F_{bc})=0$.
Calculating this expression, we find 
\be
\nabla_c(K^b\, F_{ba}) - \nabla_a(K^b\, F_{bc}) = -K^b\,\nabla_b\,  F_{ac} -
F_{ab}\, \nabla_c\, K^b - F_{bc}\, \nabla_a\, K^b\,,
\ee
which is nothing but the Lie derivative ${\cal L}_K\, F_{ca}$.  This
vanishes precisely by virtue of the assumption that $F$ is invariant
under the Killing symmetry.

   The above argument establishes that every Killing vector on the
base manifold lifts to one in the total bundle space.  In addition to
these Killing vectors of the isometry group $G$ of the base manifold,
there will also be the $U(1)$ Killing vector $\del/\del z$ on the
$U(1)$ fibres.  Thus the isometry group of the total bundle space will
be at least $G\times U(1)$. 

   We are interested in obtaining an explicit construction of the
Killing vectors in certain Einstein spaces that can be used for
Kaluza-Klein reduction, in order to test the consistency as described
in section 2.  In all the examples that we shall consider, the
Einstein space can be constructed as a $U(1)$ bundle over a \Ka base
manifold.  More specifically, in all cases of interest this \Ka space
will itself be a direct product of Einstein-\Ka spaces.  The
additional structure of the \Ka spaces allows us to obtain more
explicit constructions for the Killing vectors in the bundle space. 

\subsection{Killing vectors on \Ka spaces and their $U(1)$ bundle spaces}

   We begin with a review of some basic properties of Killing vectors
and \Ka spaces.  Consider a compact \Ka manifold $M$ equipped with a
positive definite metric $g_{ab}$, and a \Ka form $J_{ab}$.  We are
interested in the case where $M$ has continuous isometries, and hence
admits Killing vectors.  It follows from the defining equation
$\nabla_a\, K_b + \nabla_b\, K_a=0$ for a Killing vector that
\be
\square K_a + R_{ab} K^b=0\,.\label{kvlap}
\ee
Multiplying by $K^a$, integrating over $M$, and integrating by parts,
gives
\be
\int_{M} ( - |\nabla_a\, K_b|^2 + R_{ab} K^a K^b)=0\,,\label{camb1}
\ee
where $|\nabla_a\, K_b|^2$ means $(\nabla_a\, K_b)(\nabla^a\,
K^b)$. The metric is positive-definite, and so from (\ref{camb1}) we
deduce that for Killing vectors to exist, there must be appropriate
non-negative contributions from the Ricci-tensor term.  In fact we 
are interested in the case where $R_{ab}$ is positive definite.  

   Another consequence that follows from the positivity of the Ricci
tensor is that the first Betti number $b_1$ of the \Ka space must be
zero.  This follows from an argument precisely paralleling the one
above concerned with the possibility of the existence of Killing
vectors.  A harmonic 1-form $H_a$ satisfies the equation $-\square\,
H_a + R_{ab}\, H^b=0$.  By multiplying by $H^a$, integrating over $M$,
and integrating by parts on the first term, we see that there can be
no harmonic 1-forms if the Ricci tensor is positive definite.  Since
we shall be considering spaces that have strictly positive-definite
Ricci tensors, it follows that they will have $b_1=0$, and admit no
harmonic 1-forms.

   Consider now the vector $V^a$, constructed from the Killing
vector $K^a$ as follows:
\be
V^a \equiv J^a{}_b\, K^b\,.
\ee
Our first goal will be to show that $V^a$ can be written as the
gradient of a scalar function.  To prove this, define
\be
Q_{ab} \equiv \nabla_a\, V_b - \nabla_b\, V_a\,.
\ee
It follows that
\bea
|Q_{ab}|^2 &=& (\nabla_a\, V_b - \nabla_b\, V_a)
             (\nabla^a\, V^b - \nabla^b\, V^a)\,,\nn\\
&=& 2  (\nabla_a\, K_b)(\nabla^a\, K^b) -
             2 J^{ad}\, J^{cb}\,  (\nabla_a\, K_b)(\nabla_c\, K_d)\,.
\eea
Integrating this over $M$, and integrating by parts on each 
term, we obtain
\bea
\int_{M} |Q_{ab}|^2 &=& - 2 \int_{M} K^a\, \square K_a 
                         + 2 \int_{M}  J^{ad}\, J^{cb}\, K_b\,
\nabla_a\,\nabla_c\, K_d \,,\nn\\
&=&  -2 \int_{M}  K^a\, \square K_a + 2  \int_{M}  J^{ad}\, J^{cb}\, K_b\,
 R^e{}_{acd}\, K_e\,,\nn\\
&=& - 2 \int_{M} K^a( \square K_a + R_{ab} K^b)\,,\nn\\
&=&0\,,
\eea
where we have used the standard Killing-vector identity $\nabla_a\,
\nabla_c\, K_d = R^e{}_{acd} \, K_e$ in reaching the second line, and
the standard \Ka identity $R_{abcd} = J_c{}^e\, J_d{}^f\, R_{abef}$ in
reaching the third line.  The final result follows from using
(\ref{kvlap}).  Thus we conclude that $Q_{ab}=0$, and hence that
$V_a$, viewed as a 1-form, is closed; $dV=0$.  Locally, therefore, we
can write $V = - d\psi$.  As we discussed previously, we shall
be interested in \Ka spaces with positive-definite Ricci tensor, and
such spaces have vanishing first Betti number.  Since there are no
harmonic 1-forms in such spaces, it follows that $dV=0$ can be solved
globally by writing $V=-d\psi$.  In other words, we have the result
that on a \Ka space with vanishing first Betti number, any Killing
vector can be written as
\be
K^a = J^{ab}\, \del_b\psi\,,\label{kvscal}
\ee
for some scalar $\psi$.

   This scalar $\psi$ has a clear interpretation if we impose that our
\Ka space is also an Einstein space, $R_{ab}
= \Lambda g_{ab}$, where $\Lambda$ is the
``cosmological constant'' on $M$. Then (\ref{kvlap}) reduces to
\be
\square K_a + \Lambda \, K_a=0\,.\label{kvlap1}
\ee
It is now straightforward to see, by substituting (\ref{kvscal})
into (\ref{kvlap1}), that this scalar field is actually an
eigenfunction of the Laplacian on $M$, $-\square \psi = \lambda\,
\psi$, with
\be
\square \psi + 2\Lambda\, \psi = 0\,.\label{scalaref}
\ee
Moreover, the implication goes in the other direction as
well. In other words, if $\psi$ is an eigenfunction of the scalar
Laplacian, satisfying (\ref{scalaref}), then $K^a$ defined by
(\ref{kvscal}) is a Killing vector.  To see this, we define $P_{ab}
\equiv \nabla_a\, K_b + \nabla_b\, K_a$.  Substituting (\ref{kvscal})
into this, writing $\int |P|^2$, and then performing appropriate 
integrations by parts, we find that
\be
\int_{M} |P_{ab}|^2 = 2(\lambda-2\Lambda)\, \int_{M}
|\nabla_a\psi|^2\,.
\ee
(Again, standard results from \Ka geometry are needed in intermediate
steps.)  Thus if the scalar eigenfunction $\psi$ has eigenvalue
$\lambda = 2\Lambda$, it follows that $P_{ab}=0$ and hence that $K_a$
constructed as in (\ref{kvscal}) is a Killing vector.

    Thus we see that there is a one-to-one correspondence between
Killing vectors, and scalar eigenfunctions with eigenvalue
$\lambda=2\Lambda$, where $\Lambda$ is the cosmological constant of
the Einstein-\Ka space.

     Using (\ref{kvscal}), we can now obtain explicit expressions for
the Killing vectors on the space of the $U(1)$ bundle over an
Einstein-\Ka base space, where the curvature of the $U(1)$ connection
is taken to be proportional to the \Ka form.  Taking the Einstein-\Ka
base metric to have cosmological constant $\Lambda$ as above, and
taking the field strength of the connection $A$ on the $U(1)$ bundle
to be $F= \a\, J$, where $\a$ is a constant, it follows from
(\ref{ricci2}) that the Ricci tensor on the bundle space will be
given by
\be
\hat R_{ab} = (\Lambda-\ft12 c^2\, \a^2)\, \delta_{ab}\,,\qquad
\hat R_{zz} = \ft14 c^2\, \a^2\, D,\qquad
\hat R_{az}=0\,,
\ee
where $D$ is the dimension of the base manifold. In particular, the metric
on the $U(1)$ bundle becomes Einstein if $a$ is chosen such that
\be
\Lambda= \ft14 c^2\, \a^2\, (D+2)\,.
\ee
Substituting (\ref{kvscal}) into (\ref{u1_cond1}), we now obtain the
result that
\be
\del_a\, h = \del_a\, (\a\, \psi + K^b\, A_b)\,,
\ee
which can be integrated to give $h = \a\, \psi + K^b\, A_b$.
Thus for each Killing vector $K$ on the Einstein-\Ka base space, with
its associated scalar $\psi$ as given in (\ref{kvscal}), the
corresponding Killing vector in the $U(1)$ bundle space is
\be
\hat K = K + ( \a\, \psi + K^b\, A_b)\, \fft{\del}{\del z}\,.
\label{kvresult}
\ee

     It is worth noting at this stage that there is an elegant
expression for the Killing vector $\hat K$, viewed as a 1-form by
lowering its vector index using the metric (\ref{bundlemet}) on the
bundle space.  After doing this, we find that as a 1-form we have
\be
\hat K = -\im \, (\del-\bar\del)\psi + \a\, c^2\, \psi\, (dz-A)\,,
\label{kv1form}
\ee
where $\del$ and $\bar\del$ are the holomorphic and anti-holomorphic
exterior derivatives; $d=\del+\bar\del = d\zeta_\a\, \del_\a +
d\bar\zeta^{\bar\a}\, \del_{\bar\a}$.  

  Note that in the case of an Einstein-\Ka base space we can easily express
$\psi$ in terms of the Killing vector $K^a$, since from (\ref{kvscal})
we have $J^{ab}\, \del_a\, K_b = -\square\,\psi$, and hence from
(\ref{scalaref}) we shall have
\be
\psi = \fft{1}{2\Lambda}\, J^{ab}\, \del_a\, K_b\,.\label{psiexp}
\ee

\subsection{Killing vectors on a product of \Ka spaces and their
$U(1)$ bundles}

    In subsequent sections, we shall be interested in constructing
Killing vectors on $U(1)$ bundles over products of 2-spheres and more
generally complex projective spaces $\CP^n$.  These are
are particular examples of Einstein-\Ka spaces.  In this section, we
shall give results for the construction of Killing vectors
on $U(1)$ bundles over the direct product of $N$ Einstein-\Ka spaces
$M_i$, of real dimensions $d_i$, 
\ie $M = M_1 \times M_2 \times \cdots \times M_N$, with total real
dimension 
\be
D= \sum_{i=1}^N d_i\,.
\ee
The metric on the bundle space will be given by
\be
d\hat s^2 = c^2\, (dz - A)^2 + \sum_{i=1}^N ds_i^2\,,\label{metricxxx}
\ee
where $ds_i^2$ is the Einstein-\Ka metric on the factor $M_i$ in the
base space, with cosmological constant $\Lambda_i$.  The total
connection $A$ is equal to the sum of contributions from each factor,
$A= \sum_i A^{(i)}$.

     Since the base space is a direct product, we can choose the
natural block-diagonal basis for its Killing vectors, where there is
no mixing between the isometries of each factor in the product.  Thus
if $K^{(i)}$ is a Killing vector on $M_i$ then it is also a
Killing vector on $M$, and {\it vice versa}.  If we use $a_i$ to denote
a coordinate index on $M_i$, then this result follows by combining the Killing
equation $\nabla_{a_i}\, K_{b_i} + \nabla_{b_i}\,
K_{a_i}=0$ on $M_i$, with the fact that the $K^{(i)}$ are
covariantly constant with respect to $\nabla_{b_j}$ for $j \neq i$.

   This fact allows us to use the results of the previous section to
express any of these Killing vectors $K^{(i)}$ on $M$ as
\be
K^{a_i} = J^{a_i b_i}\, \del_{b_i}\psi^{(i)}\,,\label{kvscal_i}\,,
\ee
where $\psi^{(i)}$ is the corresponding scalar eigenfunction of the
Laplacian on $M_i$ with eigenvalue $2\Lambda_i$, and $J_{a_i b_i}$ are
the components of the \Ka form on $M_i$.  From the results in the
previous section, it then follows that the corresponding Killing
vector in the bundle space will be
\be
\hat K^{(i)} = K^{(i)} + (\a_i\, \psi^{(i)} + K^{b_i}\,
A_{b_i})\, \fft{\del}{\del z}\,,\label{kvprod}
\ee
where $A_{a_i}$ is the contribution to $A$ from the factor
$M_i$ in the base space, $A=\sum_i A^{(i)}$, and we are taking 
\be
F=dA = \sum_i \a_i\, J^{(i)}\,,
\ee
where $J^{(i)}$ is the \Ka form on $M_i$. 

    We may again obtain an elegant expression for the Killing vector
viewed as a 1-form, generalising (\ref{kv1form}):
\be
\hat K^{(i)} = -\im\, (\del-\bar\del)\, \psi^{(i)} + \a_i\, c^2\,
\psi^{(i)}\, (dz-A)\,.\label{kv1formprod}
\ee

    The period $\Delta z$ of the fibre coordinate $z$ must be
compatible with the integrals of $F$ over all 2-cycles in the base
manifold.  Specifically, we must have 
\be
\Delta z= \fft{1}{q_k}\, \int_{\Sigma_k} F\,,\label{top1}
\ee
where $q_k$ is an integer and $\Sigma_k$ is any 2-cycle in the base space.
We are taking each factor $M_i$ in the base space to be
Einstein-\Ka, with cosmological constant $\Lambda_i$, and so it
follows that the Ricci form $P^{(i)}$ in $M_i$ is given by
$P^{(i)}=\Lambda_i\, J^{(i)}$.  Since $1/(2\pi)\, P^{(i)}$ defines the
first Chern class of $M_i$, it follows that $1/(2\pi)\, \int
P^{(i)}=\,$integer, where the integral is taken over any 2-cycle in
$M_i$, whilst the integral will be zero for any 2-cycle in $M_j$ with
$j\ne i$.  If we define $k_i$ to be the greatest common divisor of 
the integers obtained by integrating $P^{(i)}$ over all possible
2-cycles in $M_i$, then it follows from (\ref{top1}) that $z$ must
have a period such that
\be
\Delta z = \fft{2\pi\, \a_i\, k_i}{\Lambda_i\, q_i}\,,\label{top2}
\ee
for all $i$, where the $q_i$ are integers.  Thus we must have
\be
\a_i = \fft{b\, \Lambda_i\, q_i}{k_i}\,,\label{top3}
\ee
where $b$ is related to the period $\Delta z$ by
$b=\Delta z/(2\pi)$, and it is a constant independent of $i$.  Since
we have also included the constant $c$ in (\ref{bundlemet}), we are
free to choose $b$ at will, to give $z$ a convenient period.  The
integers $q_i$ can be thought of as the winding numbers of the $U(1)$
bundle over each factor $M_i$ in the product base manifold. 

   Note that it one wants the total $U(1)$ bundle space to be
Einstein, with cosmological constant $\hat \Lambda$, then it follows
from (\ref{ricci2}) that we must have
\bea
\hat \Lambda &=& \Lambda_i -\ft12 c^2\, \a_i^2 \,,\qquad {\rm for\
all}\,\,\,  i\,, \label{einstein1}\\
\hat\Lambda &=& \ft14 c^2\, \sum_i d_i\, \a_i^2\,,\label{einstein2}
\eea
where $d_i$ is the dimension of the manifold $M_i$, and $\a_i$ is
given by (\ref{top3}).  To solve these equations, one can view
$\hat\Lambda$ and the winding numbers $q_i$ as freely-specifiable
quantities, with the $N$ equations (\ref{einstein1}) then being solved
for the individual cosmological constants $\Lambda_i$ of the factors
in the base space, and (\ref{einstein2}) being solved for the scale
factor $c$ in fibre direction of the metric (\ref{metricxxx}) on the
bundle space.  As we shall now show, one can always solve these
equations for the $\Lambda_i$ and $c$, for any choice of the integers
$q_i$, provided that they are not all zero.\footnote{If some of the
$q_i$ are zero, we can just separate off the corresponding Einstein
spaces $M_i$ in the base space, and prove the existence of an Einstein
metric on the bundle over the remaining base-space factors for which
all the $q_i$ are non-zero.  The product of this bundle space with the
Einstein spaces associated with the $q_i=0$ factors can clearly be
made Einstein, by appropriate choice of the $\Lambda_i$.  If {\it all}
the $q_i$ were zero the $U(1)$ bundle would be trivial and the total
$(D+1)$-dimensional space would be $S^1\times M$, which clearly cannot
be Einstein since the factors in the base space $M$ are assumed to
have strictly-positive cosmological constants.}

  To see this, we substitute (\ref{top3}) into (\ref{einstein1}), and
note that for each $i$ the equation allows a real solution for
$\Lambda_i$ only if
\be
c^2\, \a_i^2  \le \fft{\Lambda_i^2}{2 \hat\Lambda}\,.
\ee
Summing over $i$, and using (\ref{einstein2}) then gives
\be
\hat\Lambda^2 \le \ft18 \sum_i d_i\, \Lambda_i^2\,.\label{abcd}
\ee
On the other hand, combining (\ref{einstein1}) and (\ref{einstein2})
we have
\be
\hat\Lambda = \fft{1}{D+2}\, \sum_i d_i\, \Lambda_i\,.\label{abcd2}
\ee
Combining (\ref{abcd}) and (\ref{abcd2}) then gives the result
\be
(D+2)^2\, \sum_i d_i\, \Lambda_i^2 - 8 \sum_{i,j} d_i\, d_j\,
\Lambda_i \, \Lambda_j \ge 0\,.\label{crit1}
\ee
This is the criterion for the existence of a real Einstein space.
Since it is just a quadratic form in $\Lambda_i$, it
can be expressed as the condition that the $N\times N$ matrix
$M_{ij}$, defined by
\be
M_{ij} = (D+2)\, d_i\, \delta_{ij} - 8 d_i\, d_j\,,
\ee
must have non-negative eigenvalues. 

   To show this, we first note that the matrix $M_{ij}$ has
determinant given by
\be
\det(M_{ij}) = (D-2)^2\, (D+2)^{2N-2}\, \prod_i d_i\,,\label{detres}
\ee
which is strictly positive, since we may always assume
$D>2$.\footnote{The case where the total dimension $D$ of the base
space is equal to 2 can easily be disposed of in a separate
discussion.  The only possibility would be for the base space to be
$S^2$, and we already know that the $U(1)$ bundle over this is $S^3$,
which admits an Einstein metric.}

  Secondly, we note that if the dimensions $d_i$ are all taken
to be equal, $d_i=d$, then the eigenvalues of $M_{ij}$ are $(D+2)^2\,
d$ (occurring $N-1$ times) and $(D-2)^2\, d$ (occurring once).  Thus in
this special case all the eigenvalues of $M_{ij}$ are strictly
positive.  If $M_{ij}$ were to have any negative values for any valid
choice of the $d_i$, it would have to be the case that $\det(M_{ij})$ 
passed through 0 as the parameters $d_i$ were adjusted from $d_i=d$ to
these putative values of $d_i$.  However, we saw from (\ref{detres})
that the determinant is strictly positive, and so
it follows that $M_{ij}$ cannot have negative eigenvalues for any
valid choice of $d_i$.  Thus it is guaranteed that the inequality
(\ref{crit1}) is satisfied, and so a real solution to
the conditions (\ref{einstein1}) and (\ref{einstein2}) always exists.

  Although we have given an existence proof for an Einstein metric on
the bundle spaces for any choice of the winding numbers $q_i$, it is
not in general easy to solve explicitly for the cosmological constants
$\Lambda_i$ of the individual factors in the base space.  (In general,
one has to solve high-order polynomial equations.)  However, a simple
solution of (\ref{einstein1}) and (\ref{einstein2}) can always be
explicitly obtained in the special case where we choose the winding
numbers $q_i$ to be such that $q_i = k_i/\ell $, where $\ell={\rm
gcd}\, (k_i)$ is the greatest common divisor of the $k_i$.  In this
case, from (\ref{einstein1}) we see that this set of $N$ equations,
labelled by $i$, all become equivalent. Therefore, defining $\Lambda
\equiv \Lambda_i$ and $\a \equiv \a_i$, we have
\be
\Lambda = \fft{D+2}{D} \hat \Lambda \, , \label{lambda}
\ee 
and
\be
c^2\, \a^2 = \fft4{D+2} \Lambda \,, \label{alpha}
\ee
where $D$ is the total dimension of the base manifold. Combining
(\ref{top3}) and (\ref{alpha}) it follows that the parameters of the metric
satisfy the relation
\be
\Lambda \, b^2 \, c^2 = \fft{4\ell^2}{D+2}\, .
\ee
Note that since in this special case we have all the $\Lambda_i$
equal, the product of Einstein-\Ka base spaces is itself an Einstein space.
This situation with $q_i=k_i/\ell$ will be seen to be of particular 
significance in the next section, when we take the factors in the
product base space all to be complex projective spaces.  It turns out
that the Einstein spaces with $q_i=k_i/\ell$ then all admit 2 Killing spinors.

\section{Products of $\CP^{n}$ spaces, and their $U(1)$ bundles}

\subsection{Geometry of $\CP^n$, and its Killing vectors}

   We begin by reviewing the Fubini-Study construction of the
Einstein-\Ka metric on $\CP^n$.  Let $Z^A$ be complex coordinates on
$C^{n+1}$, with the flat metric
\be
ds_{2n+2}^2 = dZ^A\, d\bZ_A\,.\label{cn+1met}
\ee
We shall split the index $A$ into $A=(0,\a)$, where $1\le\a\le n$, 
and introduce inhomogeneous coordinates
$\zeta^\a = Z^\a/Z^0$, in the patch where $Z^0\ne 0$.  We make the
further definitions 
\be
Z^0=e^{\im \tau}\, |Z^0|\,,\qquad r=\sqrt{Z^A\, \bZ_A}\,,\qquad
f=1+\zeta^\a\, \bzeta^{\bar\a} \,.\label{3fdef}
\ee
Substituting into (\ref{cn+1met}), we find that the flat metric on
$C^{n+1}$ becomes
\be
ds_{2n+2}^2 = dr^2 + r^2\, d\Omega_{2n+1}^2\,,
\ee
where $d\Omega_{2n+1}^2$ is the metric on the unit sphere $S^{2n+1}$,
given by
\be
d\Omega_{2n+1}^2 = (d\tau + B)^2 + f^{-1}\, d\zeta^\a\, d\bzeta^{\bar\a} - 
  f^{-2}\, \bzeta^{\bar\a}\, \zeta^\beta\, d\zeta^\a\, d\bzeta^{\bar\beta}\,,
\label{hopf}
\ee
where
\be
B= \ft12\, \im\, f^{-1}\, (\zeta^{\a}\, d\bzeta^{\bar \a} - \bzeta^{\bar\a}\,
d\zeta^{\a})\,.\label{bdef}
\ee
The metric (\ref{hopf}) is the unit $S^{2n+1}$ described as a $U(1)$
bundle over $\CP^n$, and the last two terms are precisely the
Fubini-Study metric $d\Sigma_n^2$ on $\CP^n$:
\be
d\Sigma_{n}^2 =  f^{-1}\, d\zeta^\a\, d\bzeta^{\bar\a} - 
  f^{-2}\, \bzeta^{\bar\a}\, \zeta^\beta\, d\zeta^\a\, 
  d\bzeta^{\bar\beta}\,,\label{fubstud}
\ee
and so
\be
d\Omega_{2n+1}^2 = (d\tau + B)^2 + d\Sigma_n^2\,.\label{spherecpn}
\ee
The quantity $B$ defined in (\ref{bdef}) is a potential for the \Ka form, with
\be
J=dB = \im f^{-1}\, d\zeta^\a\wedge d\bzeta^{\bar\a} + \im\, f^{-2}\,
\bzeta^{\bar\a}\, \zeta^\b\, d\zeta^\a\wedge d\bzeta^{\bar\b}\,,\label{kform}
\ee
which is the \Ka form.  This can be written as 
$J = \im\, g_{\a\bb}\, d\zeta^\a\wedge d\bzeta^{\bar\b}$, where the metric
$g_{\a\bb}$ and its inverse $g^{\a\bb}$ are given by
\be
g_{\a\bb} = \ft12 f^{-1}\, \delta_{\a\bb} -\ft12 
f^{-2}\, \bzeta^{\bar\a}\, \zeta^\b
\,,\qquad
g^{\a\bb} = 2f\, \delta^{\a\bb} + 2 f\, \zeta^\a\, \bzeta^{\bar\b}
\,.\label{cpngmet}
\ee
The Fubini-Study metric (\ref{fubstud}) is Einstein, with
cosmological constant 
\be
\Lambda=2(n+1)\,.\label{cpncosmo}
\ee
We shall refer to the Fubini-Study metric (\ref{fubstud}) with this specific
normalisation for the cosmological constant as the ``unit $\CP^n$
metric,'' since it is the one that corresponds to the Hopf fibration
of the unit $(2n+1)$-sphere.  Note that $\CP^n$  has the isometry
group $SU(n+1)$, which can be seen from the fact that the metric
(\ref{cn+1met}) and the coordinate $r$ are both invariant under
$SU(n+1)$, acting by matrix multiplication on the column vector $Z^A$.

   Since we eventually want to be able to construct Killing vectors on 
$U(1)$ bundles over products of $\CP^n$ spaces, we need to find the
eigenfunctions of the scalar Laplacian on $\CP^n$ with eigenvalue
$2\Lambda$, as discussed in section (3.2).  In fact the construction of
all scalar eigenfunctions on $\CP^n$ is very simple.  Let $T_{A_1\cdots
A_p}{}^{B_1\cdots B_q}$ be a constant Hermitean $SU(n+1)$ tensor, which is
symmetric in the index set $\{A_1,\ldots,A_p\}$ and the index set
$\{B_1,\ldots,B_q\}$, and traceless in any contraction between an $A$
and a $B$ index.  This defines the $(p,q)$ representation of
$SU(n+1)$.  Clearly the scalar function
\be
\Phi=T_{A_1\cdots A_p}{}^{B_1\cdots B_q}\, Z^{A_1}\cdots Z^{A_p}\,
\bZ_{B_1}\cdots \bZ_{B_q}
\ee
is a zero mode of the Laplacian on $C^{n+1}$:
\be
\square_{C^{n+1}}\, \Phi = \fft{\del^2}{\del Z^A\, \del\bZ_A}\,
\Phi=0\,,\label{zeromode}
\ee
where we can write this Laplacian in terms of $r$ and the Laplacian on
the unit $S^{2n+1}$ as
\be
0=\square_{C^{n+1}}\, \Phi = \fft{1}{r^{2n+1}}\, \fft{\del}{\del r}
\Big(r^{2n+1}\, \fft{\del\Phi}{\del r}\Big) + \fft{1}{r^2}\,
\square_{S^{2n+1}}\, \Phi\,.\label{sphlap}
\ee
Note that $\Phi$ can be written as
\be
\Phi = r^{p+q}\, e^{\im\, (p-q)\, \tau}\, \Psi\,,\label{phidef}
\ee
where $\Psi$ depends only on the inhomogeneous $\CP^n$ coordinates
$\zeta^\a$.  

    It is straightforward to show from (\ref{spherecpn}) that the
components of the sphere metric $\hat g_{AB}$ and the $\CP^n$ metric
$g_{ab}$ are related by
\bea
&&\hat g_{ab} = g_{ab} + B_a\, B_b\,,\qquad \hat g_{a\tau} =
B_a\,,\qquad \hat g_{\tau\tau} = 1\,,\nn\\
&&\hat g^{ab} = g^{ab}\,,\qquad \hat g^{a\tau} = -B^a\,,\qquad 
\hat g^{\tau\tau} = 1 + B_a\, B^a\,,
\eea
where $B^a\equiv g^{ab}\, B_b$.  From this it is easily seen that the
scalar Laplacian on $S^{2n+1}$ is given by
\be
\square_{S^{2n+1}} = \Big(\nabla_a - B_a\, \fft{\del}{\del\tau}\Big)^2 
    + \fft{\del^2}{\del\tau^2}\,.\label{scpnlap}
\ee
Substituting (\ref{phidef}) into (\ref{sphlap}) and (\ref{scpnlap}), we
therefore find  that $\Psi$ is an eigenfunction on $\CP^n$, satisfying 
\be
-{\cal D}_a\, {\cal D}^a \, \Psi = 2[2p\, q\ + n\, (p+q)]\, \Psi\,,
\label{charged}
\ee
where ${\cal D}_a = \nabla_a -\im\,(p-q)\, B_a$.  This is the
Laplacian for scalar fields of charge $(p-q)$, in the
$(p,0,0,\ldots,0,q)$ representation of $SU(n+1)$.  The uncharged
scalars therefore occur in the $(p,0,0,\ldots,0,p)$ representations,
with eigenvalues $\lambda= 4p\, (p+n)$.

   In section (3.2) the Killing vectors on an Einstein-\Ka space were
constructed in terms of uncharged scalar eigenfunctions with eigenvalue
$2\Lambda$.  On $\CP^n$, the appropriate eigenfunctions are the ones
with $(p,q)=(1,1)$, since, as can be seen from (\ref{charged}), they
have eigenvalue $4(n+1)$,  which, from (\ref{cpncosmo}), is
$2\Lambda$.  They are indeed in the adjoint representation of $SU(n+1)$,
as should be since they are supposed to be in one-to-one
correspondence with the Killing vectors of $\CP^n$.

   Thus we see that the scalars $\psi$ that generate the Killing
vectors on $\CP^n$ are given by 
\be
\psi = \fft1{r^2}\, T_A{}^B\, Z^A\, \bZ_B\,,\label{cpnpsi0}
\ee
where $T_A{}^B$ is an arbitrary Hermitean traceless tensor.  From the previous
definitions, it has the following expression in terms of the
inhomogeneous coordinates on $\CP^n$:
\be
\psi = f^{-1}\, (T_0{}^0 + T_0{}^\a\, \bzeta^{\bar\a} + T_\a{}^0\, \zeta^\a
+ T_\a{}^\b\, \zeta^\a\, \bzeta^{\bar\b})\,.\label{cpnpsi}
\ee
Note that since $T_A{}^B$ is traceless, we can write $T_0{}^0=
-T_\a{}^\a$, and thus we can regard the unconstrained 
constant tensors $T_0{}^\a$, $T_\a{}^0$ and $T_\a{}^\b$ as
parameterising the set of scalars $\psi$ corresponding to the full set
of $n(n+2)$ Killing vectors of $\CP^n$.

   From the scalars $\psi$, we can readily construct the Killing
vectors using (\ref{kvscal}).  From (\ref{cpngmet}) we therefore find
that the complex components of the Killing vector associated with
$\psi$ are given by
\be
K^\a = \im \, g^{\a\bb}\, \del_{\bb}\, \psi
= \ft{\im}{2} (T_0{}^\a + T_\b{}^\a\, \zeta^\b - T_0{}^0 \, \zeta^\a - 
   T_\beta{}^0\, \zeta^\beta\, \zeta^\a)\,,\label{cpnkv}
\ee
with $K^{\bar \a}$ being the complex conjugate of $K^\a$.

    As a check on this construction of the Killing vectors from the
scalar eigenfunctions $\psi$, we may also construct them directly,
using the fact that they must correspond to infinitesimal $SU(n+1)$
transformations of the form $\delta Z^A= \im\, \ep\, T_B{}^A\, Z^B$ on
the homogeneous coordinates, where $T_B{}^A$ is again an arbitrary Hermitean
traceless tensor.  This translates into $\delta\zeta^\a = \delta
Z^\a/Z^0 - Z^\a/(Z^0)^2\, \delta Z^0$, giving
\be
\delta \zeta^\a = \im\, \ep\, (T_0{}^\a + T_\b{}^\a\, \zeta^\b - 
  T_0{}^0 \, \zeta^\a - 
   T_\beta{}^0\, \zeta^\beta\, \zeta^\a)\,,
\ee
which is in precise agreement with (\ref{cpnkv}), since Killing
vectors generate the coordinate transformations $\delta\zeta^\a=2\ep\,
K^\a$.  Of course we also need to know the explicit scalar functions
$\psi$, for the purpose of lifting the Killing vectors to the $U(1)$
bundle space.

   Note that $\CP^n$ is a space of constant holomorphic sectional
curvature, and in fact in terms of a real index notation the
orthonormal components of the Riemann tensor of the unit $\CP^n$ with
metric (\ref{fubstud}) are given by
\be
R_{abcd} = \delta_{ac}\, \delta_{bd} -\delta_{ad}\, \delta_{bc} +
J_{ac}\, J_{bd} - J_{ad}\, J_{bc} + 2J_{ab}\,
J_{cd}\,.\label{consthol}
\ee
It is sometimes useful to work with an explicit real metric for
$\CP^n$.  In Appendix A, we obtain an iterative construction for a
real metric on $\CP^n$, in terms of a metric on $\CP^{n-1}$.  

    It is now straightforward to follow the procedure described in
sections (3.2) and (3.3), to construct the $U(1)$ bundle space over an
arbitrary product of $\CP^n$ metrics.  Specifically, we take the base
manifold to be $M=M_1\times M_2\times\cdots \times M_N$, where $M_i$
is the complex projective space $\CP^{n_i}$, with real dimension $d_i=2n_i$.
We shall denote the total bundle spaces by
\be
Q_{n_1\, n_2 \cdots n_N}^{q_1\, q_2 \cdots q_N}\,,
\ee
where the integers $q_i$ are the winding numbers of the $U(1)$ bundle
over the factors $\CP^{n_i}$ in the base manifold.

\subsection{Killing spinors on $Q_{n_1\cdots n_N}^{q_1\cdots
q_N}$ spaces}

   As we discussed in section 3.3, one can always find a solution to
the conditions (\ref{einstein1}) and (\ref{einstein2}) for any choice
of the $q_i$.  A particularly simple case is when $q_i=k_i$.  In fact
in $\CP^{n_i}$ there is only one 2-cycle, and the integer $k_i$ is
therefore simply the result from integrating the first Chern class
$P_i/(2\pi)$ over this cycle, which turns out to give
\be
k_i =n_i + 1\,.
\ee

  In fact the Einstein spaces $Q_{n_1\, n_2\cdots n_N}^{q_1\,
q_2\cdots q_N}$ with $q_i=(n_i+1)/\ell$ where $\ell$ is the greatest
common divisor of the $(n_i+1)$ have a further nice feature, namely
that they all admit Killing spinors.  To show this, we note that the
Killing spinor equation 
\be
D_A\, \eta -\fft{\im}{2}\, \sqrt{\fft{\hat\Lambda}{D}}\, \Gamma_A\,
\eta =0\label{gamder}
\ee
has the integrability condition
\be
\ft14 \hat R_{ABCD}\, \Gamma^{CD}\, \eta -\fft{\hat\Lambda}{2D}\,
\Gamma_{AB}\, \eta=0\,,\label{integrability}
\ee
which is obtained by taking a commutator of the generalised
derivatives appearing in (\ref{gamder}).  From (\ref{integrability})
one can easily deduce that the metric on the total bundle space must
be Einstein, and furthermore that 
\be
 \hat C_{ABCD}\, \Gamma^{CD}\, \eta=0\,,
\ee
where $\hat C_{ABCD}$ is the Weyl tensor on the total space.  

   If for every space $\CP^{n_i}$ we take $q_i = k_i/\ell$, where
$\ell=\hbox{gcd}(k_i)$, then we can use
(\ref{lambda}) and (\ref{alpha}) to express the non-zero orthonormal
components of the Riemann tensor on the $U(1)$ bundle space as:
\bea
\hat R_{a_i b_i c_i d_i} &=& \fft{\Lambda}{d_i+2}\,
\Big(\d_{a_i c_i} \d_{b_i d_i}-\d_{a_i d_i} \d_{b_i c_i}\Big)\nn\\
&& + \Lambda \Big[\fft{1}{d_i+2}-\fft{1}{D+2}\Big]\,
(J_{a_i c_i} J_{b_i d_i} -  
J_{a_i d_i} J_{b_i c_i} +
2\,J_{a_i b_i} J_{c_i d_i} )\,, \nn \\
\hat R_{a_i b_i a_j b_j} &=& - \fft{2\Lambda}{D+2}
J_{a_i b_i} J_{a_j b_j}\,, \label{riemman}\\
\hat R_{a_i a_j b_i b_j} &=& -\fft{\Lambda}{D+2}
J_{a_i b_i} J_{a_j b_j}\,, \nn \\
\hat R_{\,0\,a_i\,0 \,b_i} &=& \fft{\Lambda}{(D+2)} \d_{a_i b_i}\, \nn
\eea
where $D=\sum_i d_i =\sum 2 n_i$ is the total dimension of the base space,
$\Lambda$ is the (universal) cosmological constant of the
$\CP^{n_i}$, and the indices $a_i$ label the coordinates on $\CP^{n_i}$.  (We are
using the expression (\ref{consthol}) for the Riemann tensor of
$\CP^n$, appropriately rescaled so that the cosmological constant is
$\Lambda$.) 

    From (\ref{riemman}) it follows that the non-zero components of
the Weyl tensor are
\bea
\hat C_{a_i b_i c_i d_i} &=& \Lambda \,
\Big[\fft{1}{d_i+2}-\fft{1}{D+2}\Big]\,
(\d_{a_i c_i} \d_{b_i d_i}-\d_{a_i d_i} \d_{b_i c_i} + \nn\\
&& J_{a_i c_i} J_{b_i d_i} - 
J_{a_i d_i} J_{b_i c_i} +
2\,J_{a_i b_i} J_{c_i d_i} )\,, \nn\\
\hat C_{a_i b_i a_j b_j} &=& - \fft{2\Lambda}{D+2} 
J_{a_i b_i} J_{a_j b_j}\,, \\
\hat C_{a_i a_j b_i b_j} &=& -\fft{\Lambda}{D+2} 
(\d_{a_i b_i} \d_{a_j b_j} + J_{a_i b_i} J_{a_j b_j})\,. \nn
\eea
The integrability conditions (\ref{integrability}) for the existence
of Killing spinors therefore become
\bea
&&\Gamma_{a_i b_i}\, \eta + J_{a_i c_i}\, J_{b_j d_j}\, \Gamma_{c_i
d_j}\, \eta=0\,,\label{kspinor1}\\
&& (D-d_i)\, (\Gamma_{a_i b_i} + J_{a_i c_i}\, J_{b_i d_i}\,
\Gamma_{c_i d_i} + J_{a_i b_i}\, J_{c_i d_i}\, \Gamma_{c_i d_i})\,
\eta \nn\\
&& -(d_i+2)\, J_{a_i b_i}\, \sum_{j\ne i} J_{c_j d_j}\, \Gamma_{c_j
d_j}\, \eta=0\,.\label{kspinor2}
\eea

   One can show that (\ref{kspinor2}) is implied by (\ref{kspinor1}),
and in fact the full set of independent conditions can be summarised
succinctly as follows.  Without loss of generality we can choose a
basis for the $\CP^{n_i}$ spaces in which the orthonormal components of
the \Ka forms are:
\be
J_{12}=J_{34}=J_{56}= \cdots =+1\,,
\ee
with all other components being either zero, or implied by
antisymmetry from the given ones.  The conditions (\ref{kspinor1}) and
(\ref{kspinor2}) can then be shown to be precisely equivalent to the
conditions
\be
\Gamma_{12}\, \eta = \Gamma_{34}\, \eta = \Gamma_{56}\, \eta = \cdots 
\Gamma_{D-1, D}\, \eta\,.\label{kspinor3}
\ee
Since $D$ is even, and the total bundle space has dimension $D+1$, it
follows that the spinors have $2^{D/2}$ components.  There are
$\ft12D-1$ equations in (\ref{kspinor3}), each of which implies a
halving of the original number of components, and so the final
conclusion is that there are always 2 Killing spinors in these bundle
spaces (real or complex, according to whether the spinors are
Majorana or not).  Special cases of this result that have appeared
previously in the literature include the $U(1)$ bundles over
$S^2\times S^2$, $S^2\times S^2 \times S^2$ and $\CP^2\times S^2$. 
We shall in general refer to all the $q_i=k_i/\ell$ Einstein spaces as
``supersymmetric''spaces, although of course their Killing spinors are
really only associated with supersymmetric compactifications in
certain low-dimensional examples.

\section{Consistency condition for Kaluza-Klein reductions}

     We saw in section 2 that in the cases of interest in
supergravity reductions, a criterion for the consistency of the Kaluza-Klein
reduction, and truncation to the massless gauge-boson sector, is that
the Killing vectors $\hat K^I$ associated with any gauge bosons that
are to be retained must satisfy the condition that
\be
Y(\hat K^I, \hat K^J) = \hat K^{I m}\, \hat K^J_m + \fft1{2m^2}\, 
(\hat\nabla^m\, \hat K^{I n})(\hat \nabla_m\, \hat K^J_n) \equiv
Y^{IJ}\label{consistency}
\ee
should be constant, independent of the coordinates $y$ of the internal
space.  Here, $m$ is the related to the cosmological constant
$\hat\Lambda$ of the internal Einstein space by $\hat\Lambda = D\,
m^2$, where the dimension of the internal space is $D+1$.  

   We begin by noting that the second term
in (\ref{consistency}) can be re-expressed more simply by using the
following identity:
\bea
\hat\square(\hat K^m\, \hat L_m) &=& \hat K^m\, \hat\square \, \hat L_m 
+ \hat L^m\, \hat\square \, \hat K_m + 
 2 (\hat\nabla^m\, \hat K^n)(\hat\nabla_m\, \hat L_n)\nn\\
&=& -2\hat\Lambda\, \hat K^m\, \hat L_m + 
 2 (\hat\nabla^m\, \hat K^n)(\hat\nabla_m\, \hat L_n)\,,
\eea
for any pair of Killing vectors $\hat K^m$ and $\hat L^m$, 
where we have made use of the fact that Killing vectors on an Einstein
space with cosmological constant $\hat \Lambda$ satisfy the equation
$\hat\square \hat K^m +\hat\Lambda\, \hat K^m=0$.  This allows us to
express the second term in (\ref{consistency}) in terms of 
$\hat K^m\, \hat L_m$:
\be
 (\hat\nabla^m\, \hat K^n)(\hat\nabla_m\, \hat L_n) = 
\ft12 \square(\hat K^m\, \hat L_m) + \hat\Lambda\,  \hat K^m\, \hat
 L_m\,.
\label{second}
\ee
Note that we just need the Laplacian $\square$ on the base space
here, since it is equal to the Laplacian $\hat\square$ in the bundle
space when acting on scalars that are independent of the fibre
coordinate $z$.  The quantity $Y(\hat K, \hat L)$ defined in
(\ref{consistency}), whose constancy is need for consistency, is
therefore expressible as
\be
Y(\hat K,\hat L) = \ft12 (D+2)\, \hat K^m\, \hat L_m +
\fft{D}{4\hat\Lambda}\, \square(\hat K^m\, \hat L_m)\,.\label{concon}
\ee
We shall refer to the criterion that $Y(\hat K,\hat L)$ in
(\ref{consistency}) be constant as ``The Consistency Condition'' for
short.

    With our results from the previous sections we are now able to
test the consistency condition in general, for any Einstein space
$\hat M$ that is constructed as a $U(1)$ bundle over a product of
complex projective base spaces.  Before doing so, we shall show that for
{\it any} sphere $S^n$, with its standard round metric, all the
$SO(n+1)$ Killing vectors satisfy the consistency condition.  This is
an important point not only for the discussion of Kaluza-Klein
reductions on spheres themselves, but also we shall need to make use
of this fact later in the section, when we examine the consistency
condition in more general cases.

   One way to prove that the full set of $SO(n+1)$ Killing vectors on
the sphere $S^n$ satisfy the consistency condition is by using the
fact that there are always Killing spinors on the sphere, equal in
number to the dimension of the spinors, that satisfy
\be
\hat\nabla_m\eta^A - \ft{\im}{2}\, m\, \Gamma_m \, \eta^A=0\,.
\label{kspinoreq}
\ee
From any pair of these, one can construct vectors $\hat K^{AB}_m =
\bar\eta^A\, \Gamma_m\, \eta^B$, which can easily be seen to satisfy
the Killing vector equation.  One can also show that all the Killing
vectors of $SO(n+1)$ are obtained by this means.  Furthermore, it follows from
(\ref{kspinoreq}) that $\hat \nabla_m\, \hat K_n^{AB} = \im\, m\, \bar \eta^A
\,\Gamma_{mn}\, \eta^B$.  It is now relatively straightforward to show, using
Fierz rearrangements, that the Killing vectors do indeed satisfy the
consistency condition.  

    There is another way of showing that the full set of Killing
vectors on the sphere satisfy the consistency condition, which is,
perhaps, a little more geometrically appealing.  We can describe the
unit sphere $S^n$ as the surface $x^A\, x^A = 1$ in $\R^{n+1}$, where
$x^A$ are Cartesian coordinates in $\R^{n+1}$.  The Killing vectors on
$S^n$ are then given by
\be
K_{AB} = x^A\, \del_B - x^B\, \del_A\,.\label{spherekv}
\ee
If we write $x^A=r\, u^A$, where the $u^A$ satisfy $u^A\, u^A=1$ and
are coordinates on the unit $S^n$, and $r^2= x^A\, x^A$, then the
metric on $\R^{n+1}$ is given by
\be
ds^2(\R^{n+1}) = dr^2 + r^2\, du^A\, du^A\,,
\ee
where $du^A\, du^A$ is the metric on the unit $S^n$.  If we denote by
$g_{AB}$ the metric on the unit $S^n$, it is clear that it is related
to the flat metric $\delta_{AB}$ on $\R^{n+1}$ by 
\be
g_{AB} = \fft1{r^2}\, \Big(\delta_{AB} - \fft{x^A\, x^B}{r^2}\Big)\,,
\ee
since this gives $g_{AB}\, dx^A\, dx^B = du^A\, du^A$.
An elementary calculation then shows that the inner product between
Killing vectors $K_{AB}$ and $K_{CD}$ given in (\ref{spherekv}), with
respect to the metric $g_{AB}$, is
\be
(K_{AB}\cdot K_{CD}) = \delta_{AC}\, u_B\, u_D + \delta_{BD}\, u_A\, u_C
-\delta_{AD}\, u_B\, u_C - \delta_{BC}\, u_A\, u_D\,.\label{k1k2prod}
\ee

    Now, the Laplacian on $\R^{n+1}$ is related to the Laplacian on
the unit $S^n$ by
\be
\square_{R^{n+1}} = \fft1{r^n}\, \fft{\del}{\del r}\Big(r^n\,
\fft{\del}{\del r}\Big) + \fft1{r^2}\, \square_{S^n}\,.\label{rn1sn}
\ee
 From (\ref{k1k2prod}), and $x^A= r\, u^A$, we shall have
\be
\square_{R^{n+1}}\Big( r^2\, (K_{AB}\cdot K_{CD})\Big) =  4(\delta_{AC}
\, \delta_{BD} - \delta_{AD}\, \delta_{BC})\,,
\ee
and hence using (\ref{rn1sn}) we obtain
\be
\square_{S^n}\, (K_{AB}\cdot K_{CD}) + 2(n+1)\, (K_{AB}\cdot K_{CD})
=   4(\delta_{AC} \, \delta_{BD} - \delta_{AD}\, \delta_{BC})\,.
\ee
Since the unit $S^n$ has cosmological constant $(n+1)$, which
corresponds to $m^2=1$ in (\ref{consistency}), we finally arrive at
the result that on the unit $S^n$
\be
Y(K_{AB},K_{BC}) = \delta_{AC}\, \delta_{BD} - \delta_{AD}\,
\delta_{BC}\,.
\ee
This shows that indeed all the $SO(n+1)$ Killing vectors on the sphere
$S^n$ satisfy the consistency condition.

   We now turn to the case where the internal manifold is a general
Einstein space that can be constructed as a $U(1)$ bundle over a
product of complex projective spaces, of the kind we have discussed in
the previous sections.  In section 3, we derived the expression
(\ref{kvprod}) for a Killing vector on the bundle space, and
(\ref{kv1formprod}) for its expression as a 1-form.  It is now
straightforward to calculate the inner product between any two Killing
vectors, which we shall need for testing the consistency condition.
Let us first establish the notation that we shall write the $U(1)$
Killing vector that generates translations along the fibres as
\be
U \equiv \fft{\del}{\del z}\,.\label{uup}
\ee
It is easily seen that written as a 1-form, this is
\be
U = c^2\, (dz-A)\,.\label{udown}
\ee
We shall use $\hat K_i$ to denote a Killing vector lifted from the
factor $M_i$ in the base manifold.  There are four different sectors
to consider in the consistency condition, namely $Y(U,U)$,
$Y(U,\hat K_i)$, $Y(\hat K_i, \hat K_j)$ (with $i\ne j)$ and $Y(\hat
K_i, \hat L_i)$ (where $\hat K_i$ and $\hat L_i$ are two Killing
vectors in the {\it same} factor $M_i$ in the base manifold).
  
    Taking $Y(U,U)$ first we see from (\ref{uup}) and (\ref{udown})
that $U^m\, U_m = c^2=$constant, and hence from
(\ref{concon}) we shall have $Y(U,U)=$constant.  So the $U(1)$
Killing vector by itself always satisfies the consistency condition.

    Next, consider $Y(U, \hat K_i)$.  From (\ref{uup}) and
(\ref{kv1formprod}) we have
\be
U^m\, \hat K^{(i)}_m = \a_i\, c^2\, \psi^{(i)}\,,
\ee
and so from (\ref{concon}) we obtain
\be
Y(U, \hat K_i) = \fft{\a_i\, c^2}{2\hat\Lambda}\, \Big[ (D+2)\,
\hat\Lambda - D\, \Lambda_i\Big]\, \psi^{(i)}\,.
\ee
Since $\psi^{(i)}$ is never constant (it satisfies $\square\,
\psi^{(i)} = -2\Lambda_i\, \psi^{(i)}$), it follows that for a Killing
vector $\hat K^{(i)}$ coming from the base to be included in a
consistent truncation as well as the $U(1)$ Killing vector $U$, the
quantity in square brackets would have to vanish, \ie
\be
C_i\equiv   (D+2)\, \hat\Lambda - D\, \Lambda_i=0\,.\label{con1}
\ee
We shall not analyse this condition extensively at this stage, since as we 
shall see later, more severe inconsistency problems generally occur in
other sectors.  We just note, however, that in view of the relation
(\ref{abcd2}), consistency in this sector would require
\be
\sum_j d_j\, \Lambda_j - D\, \Lambda_i=0\,.
\ee
In particular, this would be satisfied if all the $\Lambda_j$ were
equal, $\Lambda_j =\Lambda$, since $\sum_j d_j = D$ (this is the case
for all the spaces with $q_i= k_i/\ell$, \ie the ones that admit 2
Killing spinors).  However, we shall see below that the Killing vector
$\hat K_i$ will still run into other consistency problems in this
case.

    Moving on to the $Y(\hat K_i, \hat K_j)$ sector, where $\hat K_i$
and $\hat K_j$ come from different factors $M_i$ and $M_j$ in the base
space, we find from (\ref{kvprod}) and (\ref{kv1formprod}) that the
inner product for two such Killing vectors is
\be
\hat K^m_i\, \hat K_{m\, j} = \a_i\, \a_j\, c^2\, \psi^{(i)}\,
\psi^{(j)}\,.
\ee
Since the two functions $\psi^{(i)}$ and $\psi^{(j)}$ are assumed to
live in two different factors in the base space here, it follows that 
$\del^a\psi^{(i)}\, \del_a\psi^{(j)}=0$, and hence, substituting into
(\ref{concon}), we find
\be
Y(\hat K_i, \hat K_j) = \fft{\a_i\, \a_j\, c^2}{2\hat\Lambda}\, \Big[
(D+2)\, \hat\Lambda - D\, (\Lambda_i +\Lambda_j)\Big]\,\psi^{(i)}\,
\psi^{(j)}\,.
\ee
Again, since the $\psi^{(i)}$ and $\psi^{(j)}$ functions are always
non-constant, the only way for $Y(\hat K_i, \hat K_j)$ to be constant
would be if the quantity in square brackets vanished, namely
\be
C_{ij} \equiv (D+2)\, \hat\Lambda - D\, (\Lambda_i +\Lambda_j) =0
\,.\label{con2}
\ee
Again, without fully analysing this condition here we may note that in
the cases of principal interest with $\Lambda_k=\Lambda$ for all $k$ 
(the ``supersymmetric'' cases where there are 2 Killing spinors),
equation (\ref{abcd2}) now allows us to deduce that
\be
C_{ij} = -D\, \Lambda\,,
\ee
and so the consistency condition is not satisfied.  Thus we already see that
we could not include Killing vectors from both of two factors $M_i$ and $M_j$
in the base space, at least in the supersymmetric cases where all the
$\Lambda_k$ are equal.

   The fourth sector to consider is when two Killing vectors $\hat
K$ and $\hat L$ come from the same factor $M_i$ in the base
space.  In order to avoid an unnecessary profusion of indices, we shall
now suppress the ``$i$'' index that labels the particular factor in
the product base manifold where the two Killing vectors are living.
Thus the quantities $\hat K$, $\hat L$, 
$\psi$, $\wtd \psi$, $d$, $\a$, $\Lambda$ in the
following discussion all refer to this specific factor in the base space.

   Now, the calculation of the inner product of the gives the result
\be
\hat K^m\, \hat L_{m} = \a^2\, c^2\, \psi\, \wtd\psi 
+ \del^a\psi\, \del_a\wtd \psi\,,
\ee
where $K^a = J^{ab}\, \del_b \psi$ and 
$L^a = J^{ab}\, \del_b \wtd\psi$.  Substituting into $Y$
defined in (\ref{concon}), we now find
\bea
&&Y(\hat K, \hat L) = \nn\\
&&\fft1{2\hat\Lambda}\, \Big\{ 
\a^2\, c^2\, [(D+2)\, \hat \Lambda - 2 D \Lambda]\, 
\psi\,  \wtd\psi + [(D+2)\, \hat\Lambda + D\, (\a^2\,
c^2 -\Lambda)]\, \del^a\psi\, \del_a\wtd\psi\nn\\
&& \qquad + D\, (\nabla^a\nabla^b\psi)(\nabla_a\nabla_b\wtd\psi)
\Big\}\,.\label{fourth}
\eea

     This equation can be simplified  considerably, as follows.
We may invoke the fact that if we consider the case where
the base manifold has just a single factor $M_i=\CP^{n_i}$, then the
corresponding bundle space, with its Einstein metric, is the standard
round metric on the sphere $S^{2n_i+1}$.  Furthermore, we know that in
this case all the Killing vectors on $S^{2n_i+1}$ satisfy the
consistency condition, as we discussed earlier.  This, therefore,
allows us to deduce that the scalars $\psi$ and $\wtd\psi$
must satisfy equations such that (\ref{fourth}) is constant when we
take just the single factor $M_i$ in the base space.  In this case we
shall have $D=d$ (the dimension of the single space $M_i$).
Substituting into (\ref{fourth}), we then learn that
\be
-\fft{4\Lambda^2}{d+2}\, \psi\, \wtd\psi + 
\fft{4\Lambda}{d+2}\,  \del^a\psi\, \del_a\wtd\psi
+  (\nabla^a\nabla^b\psi)(\nabla_a\nabla_b\wtd\psi)
\label{fconst}
\ee
must be a constant, for any choice of $\psi$ and $
\wtd\psi$ on $M_i$.  This result\footnote{One can also prove this
result directly, as follows.  We know that any Killing vector $K^a$
satisfies $\nabla_a\, \nabla_b\, K_c = R^d{}_{abc}\, K_d$.  Since we
have $K^a= J^{ab}\, \del_b\, \psi$ here, and furthermore the Riemann
tensor on $\CP^n$ is given by (\ref{consthol}), we can conclude, after
rescaling to cosmological constant $\Lambda$ on $\CP^n$, that
\be
\nabla_a\, \nabla_b\, \nabla_c\, \psi = \fft{\Lambda}{d+2}\,
\Big[J_{ab}\, J_{cd}\, \del^d\, \psi + J_{ac}\, J_{bd}\, \del^d\, \psi
- g_{ab}\, \del_c\, \psi - g_{ac}\, \del_b\, \psi - 2g_{bc}\, \del_a\,
\psi\Big]\,,\label{3derivs}
\ee
where $d=2n$.  After some simple further manipulations, the constancy of
(\ref{fconst}) follows.}  for the
eigenfunctions $\psi$ on $\CP^n$ that they satisfy the condition that
(\ref{fconst}) is constant for any pair of such eigenfunctions.} can
now be fed back into (\ref{fourth}) in the cases that really interest
us, namely when there is more than one factor in the product base
manifold.  Specifically, we can use (\ref{fconst}) in order to
eliminate the
$(\nabla^a\nabla^b\psi)(\nabla_a\nabla_b\wtd\psi)$ terms
in (\ref{fourth}).  Thus, we can deduce that consistency in this
sector will be achieved only if
\be
Q \equiv  \del^a\psi\, \del_a\wtd\psi -\beta\, 
\psi\, \wtd\psi
\ee
is constant, where the constant $\beta$ is given by
\be
\beta = \fft{\fft{4D\, \Lambda^2}{d+2} - \a^2\, c^2\, [2D\,
\Lambda - (D+2)\, \hat\Lambda]}{
\fft{4 D\, \Lambda}{d+2} - (D+2)\, \hat\Lambda - D(\a^2\, c^2
 -\Lambda)} \,.\label{betadef}
\ee
Using (\ref{einstein1}) and (\ref{einstein2}), this can be rewritten
as
\be
\beta = \fft{4D\,\Lambda^2 - 2(\Lambda-\hat\Lambda)[2D\Lambda -(D+2)\, 
\hat \Lambda](d+2)}{4D\, \Lambda -
(d+2)[(D+2)\, \hat \Lambda + D(\Lambda-2\hat\Lambda)]}\,.\label{betadef2}
\ee

    It is easiest to analyse this condition in the case where the
Killing vector $\hat L$ is taken to be the same as $\hat K$, since
if we can show that $Y(\hat K, \hat K)$ is not a constant, then
that will show that no Killing vector from the base space can be
retained in a consistent truncation.  Let us therefore just consider
one scalar eigenfunction $\psi$, with
$\wtd \psi=\psi$.   Thus we wish to study whether the
quantity
\be
Q\equiv \del^a\psi\, \del_a\psi - \beta\, \psi^2\label{qdefn}
\ee
can be constant.  If $Q$ is constant then $\nabla_a\, Q$ will be zero,
and so we can follow the familiar strategy of integrating $(\nabla_a\, Q)^2$
over the factor $M_i=\CP^{n_i}$ in the product base manifold, where the
scalar eigenfunction $\psi$ resides.  If we can show that this integral
is positive, then it will establish that $Q$ is not constant, and
hence that the gauge boson associated to the corresponding Killing
vector cannot be retained in a consistent Kaluza-Klein reduction.

    Using integrations by parts, and the equation $\square\, \psi
=-2\Lambda\, \psi$, repeatedly, we can establish the following results:
\bea
\int \psi^2\, |\nabla\psi|^2 &=& \ft23\Lambda\, \int\psi^4\,,\nn\\
\int \nabla_a|\nabla\psi|^2\, \nabla^a(\psi^2) &=&
\ft83\,\Lambda^2\, \int\psi^4 - 2 \int|\nabla\psi|^4\,,\nn\\
\int \nabla_a|\nabla\psi|^2\, \nabla^a|\nabla\psi|^2 &=& \ft83\,
\Lambda^3\, \int \psi^4 -\fft{2(d-2)}{d+2}\, \Lambda\,
\int|\nabla\psi|^4\,,
\eea
where $|\nabla\psi|^2\equiv \nabla_a\psi\, \nabla^a\psi$ and
$|\nabla\psi|^4 \equiv (|\nabla\psi|^2)^2$.  (We have used the
relation (\ref{3derivs}) in obtaining the last of these three equations.)
Using these results, we find that
\be
\int |\nabla Q|^2 = \ft83 \Lambda\, (\beta -\Lambda)^2\, \int \psi^4 +
2\Big( 2\beta - \fft{d-2}{d+2}\, \Lambda\Big)\, \int |\nabla\psi|^4\,.
\label{dq2}
\ee

    Using this, it is possible to show that, except for ``trivial''
cases that we shall discuss below, the quantity $Q$ can never be
constant for any of the eigenfunctions $\psi$ associated with the 
Killing vectors of the $SU(n_i+1)$ factors in the isometry group of
the bundle space.  We shall first discuss the ``supersymmetric''
cases, where the winding numbers $q_i$ satisfy $q_i=k_i/\ell$, since the
proof is very simple in these cases, and furthermore they are the
examples of principal physical interest.  After that, we shall present
a complete analysis for all possible choices of winding numbers.

   As we saw in section 3, when $q_i=k_i/\ell$ the cosmological constants
of all the $\CP^{n_i}$ factors in the base space are equal, as are the
constants $\a_i$; they are given by (\ref{lambda}) and (\ref{alpha}).
Substituting these into (\ref{betadef}) we find $\beta=\Lambda$, and so
(\ref{dq2}) gives
\be
\int |\nabla Q|^2 = \fft{2\Lambda(d+6)}{(d+2)}\, \int
|\nabla\psi|^4\,.
\ee
The right-hand side is manifestly positive, and so the result that $Q$
cannot be constant follows.  

   For the general (non-supersymmetric) case with arbitrary winding numbers 
$q_i$, consider first the situation when the factor $M_i$ in the base space
where $\psi$ resides is $\CP^1$.  In this particular case, because
$\CP^1$ is the sphere $S^2$, it follows that the three eigenfunctions
$\psi$ that generate the $SO(3)$ Killing vectors actually satisfy the
equation $\nabla_a\,\nabla_b\, \psi = -\Lambda\, g_{ab}\, \psi$, and
from this it follows that on $\CP^1$ we have
\be
\int |\nabla\psi|^4 = \ft83\Lambda^2\, \int \psi^4\,.
\ee
Substituting this into (\ref{dq2}) gives
\be
\int |\nabla Q|^2 = \ft83 \Lambda\, (\beta +\Lambda)^2\, \int \psi^4\,.
\label{dq2s2}
\ee
Thus we see that in this case it must be that $Q$ is constant if and
only if $\beta=-\Lambda$.  It is easy to see from the equations
(\ref{top3}), (\ref{einstein1}), (\ref{einstein2}) and (\ref{betadef}) 
that this can happen only in the extreme case where the fibres in the
$U(1)$ bundle have a non-zero winding number {\it only} over the $S^2$
factor in the base space where $\psi$ resides.  But in this extreme
case the total space is simply the direct product of 
$S^3$ times the remaining $\CP^{n_i}$ factors in 
the base.  Not surprisingly, since $S^3$ is a group manifold, it
has  Killing vectors for which the associated quantity
$Q$ will be constant.  (Since any given Killing vector is associated
with a left-translation or right-translation under $SU(2)$.)  Aside
from this extreme case, which is certainly not the one of interest to us
in this paper, we see that $Q$ can never be constant.

    Next, consider the case where the eigenfunction $\psi$ lives in a
$\CP^2$ factor in the base space. It is
necessary, again, to determine the relation between $\int
|\nabla\psi|^4$ and $\int\psi^4$.  Clearly this will be of the form
\be
\int|\nabla\psi|^4 = c\, \Lambda^2\, \int\psi^4\,,\label{dpsipsi}
\ee
where $c$ is a pure (dimensionless) number.  It is evident from the
expressions(\ref{cpnpsi0}) or (\ref{cpnpsi}) for $\psi$ that the two
integrals on $\CP^2$ must be expressible in terms of $SU(3)$-invariant
quartic polynomials built from the traceless Hermitean tensor
$T_A{}^B$. Since there is no independent fourth-order Casimir for
$SU(3)$, it must be that both integrals in (\ref{dpsipsi}) for $\CP^2$
are pure numbers times $(T_A{}^B\, T_B{}^A)^2$, the numbers being
independent of the choice of $T_A{}^B$.  Thus the constant $c$ can be
determined by evaluating the two sides of (\ref{dpsipsi}) for any
convenient choice of eigenfunction $\psi$.  From (\ref{cpnpsi}), a
simple choice is to take the $\psi$ corresponding to $T_\a{}^\beta =
\delta_\a^\beta$, which implies $T_0{}^0=-2$, with all other
components of $T_A{}^B$ zero.  This gives
\be
\psi = 1- 3\, f^{-1}\,,
\ee
where $f$ is given in (\ref{3fdef}).  It is easy to substitute this
into (\ref{dpsipsi}), leading to the result that
\be
c = 2\,.
\ee
Finally, using this result in (\ref{dq2}), with $n=\ft12 d$, we arrive
at the following:
\be
\int |\nabla Q|^2 = \ft83\, \Lambda\, \Big[(\beta +\ft12
\Lambda)^2 + \ft14\Lambda^2\Big]\, \int\psi^4\,,
\ee
which shows that $Q$ can never be constant in this case.

   Finally, we can consider the general case where $\psi$ lives in a
$\CP^n$ factor in the base space.  Now the calculation is a little more
involved, since the ratio of $\int\nabla\psi|^4$ to $\int\psi^4$
depends on the specific choice of eigenfunction $\psi$, when $n\ge 3$.
In order to achieve the best chance of having $\int\nabla Q|^2$ be
zero, one wants the ratio of $\int|\nabla\psi|^4$ to $\int|\psi|^4$ to
be as large as possible, since then the (possibly negative) second
term on the right-hand side of (\ref{dq2}) has the best chance to
outweigh the always-positive contribution from the first term on the
right-hand side.  In the Appendix we present some calculations that
provide a determination of the largest value of this ratio; see
(\ref{intrat}) and (\ref{cmax}).  Thus from (\ref{dq2}) we find that
when $n$ is odd, we shall have
\be
\int |\nabla Q|^2 \ge \ft83\Lambda\, \Big(\beta +
\fft{2 \Lambda}{n+1}\Big)^2 \, \int\psi^4\,,\label{nodd}
\ee
whilst when $n$ is even we instead find
\be
\int |\nabla Q|^2 \ge \ft83\Lambda\, \Big[ \Big(\beta +
\fft{2n\, \Lambda}{n^2+n+2}\Big)^2 + \fft{4(n+2)}{(n^2+n+2)^2}
\Big]\, \int\psi^4\,,\label{neven}
\ee

    From these results we see that $Q$ can never be constant when $n$
is even.  When $n$ is odd instead, we see that $Q$ can be constant if
and only if 
\be
\beta = - \fft{2\Lambda}{n+1}\,.\label{qconst}
\ee
Now from (\ref{einstein1}) and (\ref{einstein2}) it immediately
follows that if we define $x\equiv \hat\Lambda/\Lambda$, then 
\be
\fft{d}{d+2} \le x \le 1\,.\label{xrange}
\ee
The lower limit is saturated if the $U(1)$ fibres wind only over the
chosen $\CP^n$ factor in the base space, whilst the upper limit is
saturated if instead the $U(1)$ fibres have zero winding number over
the chosen $\CP^n$ factor.  Using (\ref{betadef2}), we find that
\be
\beta + \fft{4\Lambda}{d+2} = \Lambda\, \fft{-2(D+2)(d+2)^2\, x^2 +
2(d+2)(3D\, d + 8D + 2d)\, x - 4d\,D\,  (d+4)}{(d+2)[(D-2)(d+2)\, x
-D\, (d-2)]}\,.
\ee
The denominator is positive for all $x$ in the interval
(\ref{xrange}), and the numerator has no extremum in this interval.
It then follows that we shall have
\be
\beta \ge -\fft{4\Lambda}{d+2}\,,
\ee
with equality being achieved only if $x=d/(d+2)$.  Since $d=2n$, we
conclude from this and (\ref{qconst}) that $Q$ can be constant only in the
extreme case where the $U(1)$ fibres wind purely over the $\CP^n$ factor
in the base space in which the eigenfunction $\psi$ resides.  

    The reason for the occurrence of these exceptional cases where $Q$
can be constant is the following.  When $n$ is odd, say $n=2q+1$, and
the fibres wind only over the $\CP^{2q+1}$ factor in the base manifold,
the total space is the direct product of $S^{4q+3}$ with the
other $\CP^{n_i}$ factors in the base space.  Now the sphere $S^{4q+3}$
can be described as an $SU(2)$ bundle over the quaternionic projective
space $\HP^q$.  Consequently, an $SU(2)$ subgroup of the $SO(4q+4)$
isometry group of the sphere corresponds to left translations by
$SU(2)$ on the $SU(2)$ fibres, and therefore the associated $SU(2)$ Killing
vectors $K^I$ will necessarily have the property that $K^I\cdot
K^J=$constant, and so they will be associated with eigenfunctions
$\psi$ on $\CP^{2q+1}$ that satisfy the condition $Q=$constant.  It is
these Killing vectors that are being ``detected'' by the saturation of
the bound (\ref{nodd}). 

    These exceptional cases are higher-dimensional generalisations of
the exception arising for $n=1$, with the fibres winding only over the
$\CP^1$ factor to give $S^3$, which we discussed previously.  Again
they are ``trivial,'' from the point of view of our analysis of
compactifications, since we are not particularly interested in cases
where the internal space is a direct product of a sphere $S^{4q+3}$
with a \Ka space.  Nonetheless, it is reassuring to find that our
rather intricate general analysis has indeed, as it should, detected
these slightly obscure exceptions to the general rule.

   With these results, we have proved that the non-abelian Killing
vectors on the $U(1)$ bundle spaces over any product of $\CP^{n_i}$
factors in the base space will never satisfy the consistency
requirement that $Y^{IJ}$ in (\ref{consistency}) is a
constant.\footnote{Except in the previously-discussed trivial cases of
$SU(2)$ Killing vectors in the $S^{4q+3}$ factors in a bundle space
where the fibres wind only over a $\CP^{2q+1}$ base-space factor.}  This
means that the associated Kaluza-Klein Yang-Mills fields associated
with the non-abelian part of the symmetry group cannot be consistently
retained in a massless truncation.  In particular, this proves that of
the $U(1)\times SU(2)\times SU(2)$ Yang-Mills fields in the $Q(1,1)$
compactification of the type IIB theory to $D=5$, the $SU(2)\times
SU(2)$ fields cannot be retained in a consistent massless truncation.

\section{Conclusions}

    In this paper, we have studied a necessary condition for the
occurrence of a consistent Kaluza-Klein reduction on an internal
Einstein manifold, in which all the Yang-Mills fields associated with
the isometry group of the compactifying space are retained in a
massless truncation.  This condition, that the quantitiy $Y^{IJ}$
defined in (\ref{consistency}) should be constant, is of rather
general relevance in all the known non-trivial consistent Kaluza-Klein
reductions.  In particular, this consistency criterion is satsified by
all the Killing vectors on a sphere, of arbitrary dimension.  Our
principal goal in this paper has been to show that the consistency
criterion is never satisfied by the non-abelian $SU(n_i+1)$ Killing
vectors in the isometry groups of the spaces $Q_{n_1\cdots
n_N}^{q_1\cdots q_N}$, which are defined as $U(1)$ bundles over
the product $\prod_i \CP^{n_i}$ of complex-projective spaces
$\CP^{n_i}$, with winding numbers $q_i$.  In particular, this shows
that space $Q_{11}^{11}$ (sometimes known as $T^{11}$), the $U(1)$
bundle over $S^2\times S^2$, does not allow a consistent Kaluza-Klein
reduction of type IIB supergravity in which the non-abelian Yang-Mills
fields of its $SU(2)\times SU(2)\times U(1)$ isometry group are
retained in a massless truncation.  Likewise, the compactifications of
$D=11$ supergravity on the $U(1)$ bundles over $S^2\times S^2\times
S^2$ and over $\CP^2\times S^2$ do not allow the retention of the
corresponding non-abelian Yang-Mills fields in massless truncations.
These facts will be of relevance in the AdS/CFT correspondence
\cite{malda,gkp,wit}, where it should turn out that certain
correlation functions involving products of single massive operators
with massless ones will correspondingly be non-zero in these cases
(see, for example, \cite{dhoker}).  

    We have set our proof of the inconsistency of the full massless
truncations in these cases in a more general context, in which we
show in general that the non-abelian Killing vectors on the bundle
spaces $Q_{n_1\cdots n_N}^{q_1\cdots q_N}$ do not satisfy the
consistency criterion that all Killing vectors on all spheres
satisfy.  In order to show this, we have made a detailed analysis that
should also be of more general utility.  In particular, we 
studied the lifting of Killing vectors from an arbitrary base manifold
to a $U(1)$ bundle over the base, and then we specialised to the case
where the base is a \Ka-Einstein space, or a product of \Ka-Einstein
spaces.  In such cases, more complete results can be obtained, based
on the fact that any Killing vector in the base can be expressed in
terms of a certain eigenfunction of the scalar Laplacian.  

    We then turned to the cases of principal interest, where the base
space is the product of complex projective spaces $\CP^{n_i}$.  We
made a study of the Fubini-Study metrics, and in an appendix we
obtained a rather useful iterative construction for real metrics on
$CP^n$.  We showed that all of the bundle spaces $Q_{n_1\cdots
n_N}^{q_1\cdots q_N}$ can be given Einstein metrics, provided only
that all the winding numbers $q_i$ do not vanish.  We also showed that
in the special case where $q_i=(n_i+1)/\ell$, where $\ell$ is the
greatest common divisor of the $(n_i+1)$, the Einstein spaces all
admit two Killing spinors.  These cases, for $Q_{11}^{11}$,
$Q_{111}^{111}$ and $Q_{21}^{32}$, are the ones of principal interest
in the context of supergravity compactifications, since they imply the
existence of unbroken supersymmetries.

    We showed also that the question of whether the non-abelian
Killing vectors of the $U(1)\times \prod_i SU(n_1+1)$ isometry group
of $Q_{n_1\cdots n_N}^{q_1\cdots q_N}$ satisfy the consistency
criterion in (\ref{consistency}) can be reduced to the question of
whether the scalar eigenfunctions on $CP^n$ that are related to its
Killing vectors satisfy certain integral bounds.  We studied these
bounds in detail, and used these to obtain our proofs of the
inconsistency of the Kaluza-Klein reductions.

\section*{Acknowledgements}

     We are grateful to Mirjam Cveti\v{c}, Hong L\"u, Arta Sadrzadeh,
Kelly Stelle and Tuan Tran for helpful discussions.  C.N.P. thanks the
Caltech-USC Center for Theoretical Physics for hospitality during the
completion of this work.

\bigskip\bigskip
\noindent{\Large Appendices}
\appendix
\section{An iterative construction of $\CP^n$}

   On occasion, it is helpful to have a real expression for the
Fubini-Study metric on $\CP^n$ available.  This is easily done for
low-dimensional examples by making specific adapted coordinate choices
(see, for example, \cite{gibpop} for an explicit real metric on
$\CP^2$).  In general, we can give an elegant iterative construction
for the metric on $\CP^n$ in terms of the metric on $\CP^{n-1}$.  

    We take as our starting point the standard Fubini-Study metric
(\ref{fubstud}) on $\CP^n$, and write the inhomogeneous coordinates
$\zeta^\a$ as
\be
\zeta^\a = \tan\xi\, u^\a\,,\qquad \hbox{with}\qquad u^\a\, \bar
u^{\bar\a}= 1\,.
\ee
With this coordinate redefinition the $\CP^n$ metric (\ref{fubstud})
becomes
\be
d\Sigma_n^2 = d\xi^2 + \sin^2\xi\, du^\a\, d\bar u^{\bar\a} -
\sin^4\xi\, |\bar u^{\bar\a}\, d u^\a|^2\,.\label{fubstud3}
\ee
Noting that the $n$ quantities $u^\a$ are themselves complex
coordinates on $C^n$, subject to the constraint $ u^\a\, \bar
u^{\bar\a}= 1$, we can follow the same strategy as in the original
$\CP^n$ construction, by introducing $(n-1)$ inhomogeneous coordinates
$v^i$, with $1\le i\le n-1$, defined by
\be
v^i = \fft{u^i}{u^n}\,,
\ee
where $u^n$ here denotes the $n$'th of the coordinates $u^\a$.  In
addition, we define
\be
u^n = |u^n|\, e^{\im\, \td \tau}\,.
\ee

    After a little calculation, we see that the metric (\ref{fubstud})
on $\CP^n$ now takes the form
\be
d\Sigma_n^2 = d\xi^2 + \sin^2\xi\, \cos^2\xi \, (d\td\tau + \wtd B)^2 +
\sin^2\xi\, d\Sigma_{n-1}^2\,,\label{fubstudit}
\ee
where $d\Sigma_{n-1}^2$ is the Fubini-Study metric on the unit
$\CP^{n-1}$, and $\wtd B$ is a potential for the \Ka form of $\CP^{n-1}$:
\bea
d\Sigma_{n-1}^2 &=& \td f^{-1}\, dv^i\, d\bar v^{\bar i}\, - \td
f^{-2}\, |\bar v^{\bar i}\, dv^i|^2\,,\qquad 
\td f = 1 + v^i\, \bar v^{\bar i}\,,\nn\\
\wtd B &=& \ft12 \, \im\, \td f^{-1}\, (v^i\, d\bar v^{\bar i} - \bar
v^{\bar i}\, dv^i)\,.
\eea
Thus (\ref{fubstudit}) gives us an iterative construction of the
Fubini-Study metric on the unit $\CP^n$ in terms of the Fubini-Study
metric on the unit $\CP^{n-1}$.  (In fact the metric in
$\CP^2$ obtained in \cite{gibpop} is precisely of this form, with the
metric on $\CP^1$ being the standard metric on the 2-sphere.)  
Note that the potential $B$ for
$\CP^n$, appearing in (\ref{kform}), is given in terms of the analogous
potential $\wtd B$ for $\CP^{n-1}$ by
\be
B=\sin^2\xi\, (d\td \tau + \wtd B)\,.
\ee
The function $f$ appearing in (\ref{3fdef}) is given by
\be
f= \sec^2\xi\,.
\ee

\section{Inequalities on $\CP^n$}

    In section 5, we show that the gauge boson associated with any
Killing vector on a factor $\CP^n$ in the base manifold whose
associated scalar harmonic $\psi$ has a $Q$, as defined in
(\ref{qdefn}), that is non-constant, cannot be retained in a
consistent massless Kaluza-Klein reduction.  In this appendix we
derive some inequalities involving the integrals $\int\psi^4$ and
$\int|\nabla\psi|^4$ appearing in (\ref{dq2}), which are used in the
calculations in section 5.

     In terms of the construction (\ref{cpnpsi0}) or (\ref{cpnpsi})
for the eigenfunctions $\psi$, it is clear that the integrals
$\int\psi^4$ and $\int|\nabla\psi|^4$ must necessarily give rise to
quartic $SU(n+1)$ invariants built from the traceless Hermitean tensor
$T_A{}^B$.  If we define
\be
I_2\equiv T_A{}^B\, T_B{}^A\,,\qquad 
I_4 \equiv  T_A{}^B\, T_B{}^C\, T_C{}^D\, T_D{}^A\,,\label{i2i4}
\ee
it follows therefore that on $\CP^n$ we must have 
\be
\int\psi^4 = a\, (I_2)^2 + b\, I_4\,,\qquad \int|\nabla\psi|^4 = 
\td a\, (I_2)^2 + \td b\,I_4\,,\label{integrals}
\ee
for pure numbers $a$, $b$, $\td a$ and $\td b$ that are dependent only
on the value of $n$.  In order to determine these constants, it
suffices to consider just two special cases of eigenfunctions $\psi$
that have different values for the ratio $I_4/(I_2)^2$.

    A convenient choice for the two eigenfunctions $\psi_1$ and
$\psi_2$ is as follows.  For $\psi_1$, we take
$T_\a{}^\beta=\delta_\a^\beta$, $T_0{}^0=-n$, with all other
components of $T_A{}_B$ vanishing.  For $\psi_2$, we take instead
$T_n{}^0=T_0{}^n=\ft12$, with all other components vanishing.  (Here ``$n$''
indicates that $\a$ takes the value $\a=n$.)  For these two special
cases the invariants $I_2$ and $I_4$ are given by:
\bea
\psi_1:&&  I_2 = n(n+1)\,,\qquad I_4 = n(n^3+1)\,,\nn\\
\psi_2:&& I_2 = \ft12\,,\qquad\quad \quad I_4=\ft18\,.
\eea
Thus when $n\ge2$, we see that $I_4/(I_2)^2$ is different for the two
eigenfunctions, and so by evaluating the integrals in
(\ref{integrals}), we shall be able to determine $a$, $b$, $\td a$ and
$\td b$.  

    In order to evaluate the integrals, it is convenient to make use
of the iterative construction of $\CP^n$ metrics that we obtained in
Appendix A.  Specifically, we iterate twice, to give
\be
d\Sigma_n^2 = d\xi^2 + \sin^2\xi\, \cos^2\xi\, (d\td\tau + \wtd B)^2 +
\sin^2\xi\,\Big( d\lambda^2 +\sin^2\lambda\, \cos^2\lambda\, (dz+ C)^2
+\sin^2\lambda\, d\Sigma_{n-2}^2\Big)\,.\label{cpncpnm2}
\ee
(Our notation should be self-evident, by comparing with the construction in
Appendix A.)  The two eigenfunctions $\psi_1$ and $\psi_2$ can then be
seen to be given by
\be
\psi_1 = 1 - (n+1)\, \cos^2\xi\,,\qquad \psi_2 = \sin\xi\, \cos\xi\,
\cos\lambda\, \cos\td\tau\,.
\ee
Other relevant points are that the determinant of the metric
(\ref{cpncpnm2}) is given by
\be
\sqrt{g_n} = (\sin\xi)^{2n-1}\, \cos\xi\, (\sin\lambda)^{2n-3}\,
\cos\lambda\, \sqrt{g_{n-2}}\,,
\ee
where $g_{n-2}$ is the determinant of the metric $d\Sigma_{n-2}^2$ 
on $\CP^{n-2}$.  Furthermore, the relevant components of the inverse
metric are given by
\be
g^{\xi\xi} = 1\,,\qquad g^{\lambda\lambda}=\fft1{\sin^2\xi}\,,\qquad
g^{\td\tau\td\tau} = \fft{\sec^2\xi+ \tan^2\lambda}{\sin^2\xi}\,.
\ee
For functions $\phi$ of $\xi$, $\lambda$ and $\td\tau$ only, we have
\be
|\nabla\phi|^2 = \Big(\fft{\del\phi}{\del\xi}\Big)^2 +
\fft1{\sin^2\xi}\,  \Big(\fft{\del\phi}{\del\lambda}\Big)^2 +
 \fft{\sec^2\xi+ \tan^2\lambda}{\sin^2\xi}\,  
\Big(\fft{\del\phi}{\del\td\tau}\Big)^2\,.
\ee
As a final preliminary, we note from (\ref{hopf}) that since the unit
sphere $S^{2n+1}$ has volume $\Omega_{2n+1}= 2\pi^{n+1}/\Gamma(n+1)$,
it follows that the unit $\CP^n$ has volume $\Sigma_n$ given by
\be
\Sigma_n = \fft{\pi^n}{\Gamma(n+1)}\,.
\ee

    It is now straightforward to evaluate all the necessary integrals,
and thus to determine the constants $a$, $b$, $\td a$ and $\td b$
appearing in (\ref{integrals}).  We find that for the unit $\CP^n$,
with $n\ge 2$, we shall have
\bea
\int\psi^4 &=& \fft{3\pi^{n-1}}{2\Gamma(n+5)}\, \Big[ (I_2)^2 + 2
I_4\Big]\,,\nn\\
\int|\nabla\psi|^4 &=& 
\fft{8\pi^{n-1}}{\Gamma(n+5)}\, 
\Big[ (n^2+5n+7)\, (I_2)^2 +
(n+1)(n+2)\, I_4\Big]\,.\label{integralsi2i4}
\eea
   
    For the discussion in section 5, it turns out that we need to know
the largest possible value that the ratio $(\int
|\nabla\psi|^4)/(\int\psi^4)$ can attain.  It is easy to see from
(\ref{integralsi2i4}) that this
will occur for a tensor $T_A{}^B$ that gives the smallest possible
value of $I_4/(I_2)^2$.  To determine this value, let the (real) eigenvalues
of $T_A{}^B$ be $\lambda_A$.  Tracelessness implies that $\sum_A
\lambda_A=0$.  If we solve for the eigenvalue $\lambda_0$ in terms of
the $\lambda_\a$ for $1\le\a\le n$, we shall therefore have
\be
I_2 = \sum_\a \lambda_\a^2 +(\sum_\a \lambda_\a)^2\,,\qquad
I_4 = \sum_\a \lambda_\a^4 +(\sum_\a \lambda_\a)^4\,,
\ee
and so the ratio $R\equiv I_4/(I_2)^2$ is extremised when the
$\lambda_\a$ satisfy
\be
\lambda_\a^3\, I_2 - \lambda_\a\, I_4 + (\sum_\beta \lambda_\beta)^3\,
I_2 - \sum_\beta\lambda_\beta\, I_4=0\,.\label{derivr}
\ee

    If we suppose that two of the extremising eigenvalues, say
$\lambda_\a$ and $\lambda_\beta$, are unequal, then by subtracting
their equations (\ref{derivr}) we find that
\be 
\lambda_\a^2 + \lambda_\a\, \lambda_\beta + \lambda_\beta^2 =
\fft{I_4}{I_2}\,.\ee
If a third eigenvalue, say $\lambda_\gamma$, is unequal to both
$\lambda_\a$ and $\lambda_\beta$, then by subtractions we can see that
\be
\lambda_\a +\lambda_\beta + \lambda_\gamma=0\,.
\ee
Finally, if we suppose that a fourth eigenvalue $\lambda_\delta$ is
unequal to all of the previous three, then by subtractions we arrive
at the contradiction that $\lambda_\delta=\lambda_\a$.  Therefore
any set of $\lambda_\a$ that extremises the ratio $R$ can involve at
most three different values.

  It now becomes rather straightforward to find the extrema, and in
particular, to identify the global minima.  There are two distinct
cases, depending upon whether $n$ is even or odd.  We find that the
minimum is achieved for
\bea
{n=2q+1}:&& \lambda_0=\lambda_1=\cdots =\lambda_q=\lambda\,,\quad
\lambda_{q+1}=\lambda_{q+2}=\cdots = \lambda_{2q+1}=-\lambda\,,\\
{n=2q}:&& \lambda_0=\lambda_1=\cdots =\lambda_q=\lambda\,,\quad
\lambda_{q+1}=\lambda_{q+2}=\cdots =\lambda_{2q}=
-\fft{n+2}{n}\, \lambda\,.\nn
\eea
(Of course in each case there are symmetry-related minima
corresponding to permuting the eigenvalues.  The minimisation occurs
when the set of eigenvalues $\lambda_i$ divides into two subsets that
are as nearly as possible equal in size, within each of which all
eigenvalues are equal. This 50/50 partitioning is exact only if $n$ is odd,
since the total number of eigenvalues $\lambda_i$ is then even.)  Thus
we find that the following inequalities hold:
\bea
{n=2q+1}:&& \fft{I_4}{I_2^2} \ge \fft1{n+1}\,,\nn\\
{n=2q}:&& \fft{I_4}{I_2^2} \ge \fft{n^2+2n+4}{n(n+1)(n+2)}\,.
\eea

     Substituting into (\ref{integralsi2i4}), and reinstating the
cosmological constant $\Lambda$ by the appropriate constant rescaling,
we then find that 
\be
\int|\nabla\psi|^4 \le c_{\rm max}\, \int\psi^4\,,\label{intrat}
\ee
with
\bea
{n=2q+1}:&& c_{\rm max} = \fft{4(n+3)\, \Lambda^2}{3(n+1)}\,,\nn\\
{n=2q}:&& c_{\rm max} = 
\fft{4(n+1)(n+2)\, \Lambda^2}{3(n^2+n+2)}\,.\label{cmax}
\eea
These inequalities are used in section 5.

\end{document}